     \newcommand{\CLASSINPUTtoptextmargin}{1.25in} 
    \newcommand{\CLASSINPUTbottomtextmargin}{1.25in}
\documentclass[12pt, draftcls,onecolumn]{IEEEtran}
\usepackage{polpack}

\newcommand{\eqnsize}{\normalsize}
\usepackage{placeins}
\usepackage{accents}
 
\begin{document}
\allowdisplaybreaks[4]
\title{Content-Level Selective Offloading in Heterogeneous Networks: Multi-armed Bandit Optimization and Regret Bounds}
\author{\vspace{-0.3cm}\IEEEauthorblockN{Pol Blasco and Deniz G{\"u}nd{\"u}z} \\
\IEEEauthorblockA{\vspace{-0.3cm}Imperial College London,  UK\\ \vspace{-0.3cm}
Emails: \{p.blasco-moreno12, d.gunduz\}@imperial.ac.uk}
}
\maketitle
\vspace{-2.1cm}

\begin{abstract} \vspace{-0.5cm} 
We consider \emph{content-level selective offloading} of cellular downlink traffic to a wireless \emph{infostation} terminal which stores high data-rate content in its cache memory. Cellular users in the vicinity of the infostation can directly download the stored content from the infostation through a broadband connection (e.g., WiFi), reducing the latency and load on the cellular network. The goal of the infostation cache controller (CC) is to store the most popular content in the cache memory such that the maximum amount of traffic is offloaded to the infostation. In practice, the popularity profile of the files is not known by the CC, which observes only the instantaneous demands for those contents stored in the cache. Hence, the cache content placement is optimised based on the demand history and on the cost associated to placing each content in the cache. By refreshing the cache content at regular time intervals, the CC gradually learns  the popularity profile, while at the same time exploiting the limited cache capacity in the best way possible.  This is formulated as  a \emph{multi-armed bandit (MAB) problem with switching cost}. Several algorithms  are presented to decide on the cache content over time. The performance is measured in terms of \emph{cache efficiency}, defined as the amount of net traffic that is offloaded to the infostation. In addition to theoretical regret bounds, the proposed algorithms are analysed through numerical simulations. In particular,  the impact of system parameters, such as the number of files, number of users,  cache size, and skewness of the popularity profile, on the performance is studied numerically. It is shown that the proposed algorithms  learn the popularity profile quickly for a wide range of system parameters.  
\end{abstract}

\section{Introduction} 
Wireless data traffic represents a significant portion of the Internet traffic, and with the increasing demand for high data rate and delay intolerant applications, such as video streaming and online gaming, the growth in downlink wireless traffic is envisioned to continue in the upcoming years. A large fraction of the wireless data traffic is served from content delivery networks (CDNs), which are distributed systems of data centres  located in the core network. Content is stored in a CDN's data servers and delivered to users from the nearest location. Traditional CDNs reduce latency and alleviate congestion in the core network. However, with the increasing mobile data traffic, the access and backhaul connections of cellular networks are becoming the bottleneck for the quality-of-experience (QoE) of wireless users, and there is a growing pressure to bring the content even closer to the wireless end users.  

Pushing CDNs closer to mobile users, that is,   caching part of CDN's content at the network edge, in order to reduce delay, and alleviate congestion in the cellular backhaul has received great attention both from the academia \cite{caching:S.Borst2010,caching:Vakali2003,cahing:Maddah-Ali2014,caching:Golrezaei2012a,caching:Poularakis2013,caching:E.Bastug2014a,Blasco2014,P.Blasco2014,caching:A.Sengupta2014}, as well as the industry, with the development of storage-enabled base stations (BSs), such as CODS-AN form Saguna and  Data-at-the-Edge\texttrademark~from Altobridge.  Popular content, such as news feeds and YouTube videos, can be stored in cache memories located at wireless access points, such as BSs and small BSs (sBSs), and this content can be quickly and reliably delivered to users when requested, without consuming bandwidth in the backhaul connection.  However, the appropriate business model for the implementation of these network edge CDNs owned by the mobile network operator (MNO) raises further challenges, as it requires significant coordination and data exchange between the MNOs, traditional CDNs and content providers (CPs).

In this paper we study \emph{caching as a service}, that is, an independent third party service provider installs wireless infostation terminals, which store high data rate content (e.g., video), and serve these directly to its users through a high data-rate connection \cite{caching:infostation_Goodman1997}. These infostations may be installed and managed by CPs, such as YouTube and Netflix, in order to improve the QoE for its users at locations where cellular traffic may be congested, such as metro stations and stadiums, or by another entity which may charge the CPs or its subscribers for improving the QoE.  

Most of the existing literature on caching \cite{caching:S.Borst2010,caching:Vakali2003,cahing:Maddah-Ali2014,caching:Golrezaei2012a,caching:Poularakis2013,caching:E.Bastug2014a} assumes that files' popularity profile are known in advance, and the optimization of the cache content is carried out based on this information. Note that even in scenarios in which the content popularity is static, obtaining this information for the locality of each BS requires significant coordination between the MNOs and CPs.  We take a more practically relevant approach and consider that the popularity profiles are not known in advance. Instead, assuming stationary file popularity, we derive algorithms that learn the best caching strategy over time by observing the instantaneous demands in real time.

Since the infostation cache capacity is relatively small compared to the set of all files available in the Internet,  the likelihood that a random request from a user can be served from the cache is small, and the infostation can be locked by user requests that can not be served. To avoid this congestion and reduce the overhead, the infostation periodically broadcasts information about its cache content (i.e., a database of the stored files) to its users. When a file requested by a user is located in the cache, the request is offloaded to the infostation, otherwise it is sent to the cellular network together with all other requests.  This, we call \emph{content-level selective offloading}.

The infostation cache controller (CC) is in charge of managing the cache content. Storing new content in the cache has a cost that is related to the bandwidth consumption in the infostation backhaul link. The CC can learn/observe the popularity of each content by storing it in the cache and observing the instantaneous demands; but, at the same time, it has to take the most out of the scarce cache capacity by caching the content that it believes to be the most popular. The  objective of the CC is to find the best set of files to cache in order to maximise the traffic offloaded to the infostation  without knowing the popularity profile in advance, and by observing only the requests corresponding to the files in the cache. We model this as a multi-armed bandit (MAB) \cite{MAB:P.Auer2002} problem,  and provide several algorithms for cache content management. The main contributions of the paper can be summarised as follows:\begin{itemize}
 \item We address the optimal content placement problem in an infostation when the popularity profile of the available content is not known in advance, and placing new content into the infostation cache has a cost. 
 \item We show that this content placement problem can be formulated as an MAB-problem. Then we propose an algorithm and prove non-trivial regret bounds on its performance which hold uniformly over  time. 
 \item We propose a number of \emph{practical} algorithms that efficiently learn the popularity profile and cache the best files.  
 \item We provide extensive numerical results to study the impact of several system parameters (i.e., content popularity profile, cache size, number of users, and number of files)  on the system performance.
 \item We measure numerically the loss due to the lack of information about the popularity profile by comparing the performance of our algorithms with that of an upper bound that knows the file popularity profile. 
 \end{itemize}

The rest of the paper is structured as follows:  a survey of the relevant literature and background is presented in Section \ref{CAsec:RW}. The system model and the problem statement are presented in Section~\ref{CAsec:SM}. In Section~\ref{CAsec:CMAB} we propose an algorithm for the optimal caching problem when the popularity profile is not known and prove bounds on its performance. Several practical algorithms  are presented in Section \ref{CAsec:mab_app}. Section~\ref{CAsec:res} presents extensive numerical results and, finally, Section~\ref{CAsec:con} concludes the paper.

\section{Related Work}\label{CAsec:RW}
Content caching has been studied for wired networks and CDNs \cite{caching:S.Borst2010,caching:Vakali2003}. Recently, due to the explosion in wireless data traffic, and the introduction of sBSs with limited backhaul capacity, content caching in wireless networks has regained popularity \cite{cahing:Maddah-Ali2014,caching:Golrezaei2012a,caching:Poularakis2013,caching:E.Bastug2014a,Blasco2014,P.Blasco2014,caching:A.Sengupta2014}. 

Considering the huge number of potential content with varying sizes and popularities, an important problem is to decide which content should be cached in the limited storage space available. In  \cite{caching:S.Borst2010}, a cache cluster formed by several leaf-cache nodes and a parent-cache node is studied. The problem of optimally placing content in  the cache nodes in order to minimise the total bandwidth consumption is studied, and approximate solutions  are given for special cases. The broadcast nature of the wireless transmission is exploited in  \cite{cahing:Maddah-Ali2014} by using coded multicast transmission and storing content into end user devices. Reference  \cite{caching:Golrezaei2012a} considers a backhaul-constrained small-cell network, in which users can connect to several storage-enabled sBSs. The cache content placement that minimises the average system latency is studied and shown to be NP-hard, and approximate algorithms are given. In \cite{caching:Poularakis2013} users move randomly across a wireless network.  At each time slot, users can access a single sBS, and download only a part of a content from the sBS cache. Coded content caching in the sBSs is optimised such that the amount of users that fetch the entire content directly from the sBSs is maximised.  Outage probability and average content delivery rate in a cache-enabled sBS network is studied in \cite{caching:E.Bastug2014a}. Content placement with unknown file popularity has been studied in  \cite{Blasco2014,P.Blasco2014} and \cite{caching:A.Sengupta2014} for an infostation and a sBS network, respectively. We extend the results in \cite{Blasco2014,caching:A.Sengupta2014} and \cite{P.Blasco2014} by taking into account the cost associated to placing content into the cache and by providing a regret bound for an algorithm as well as extensive numerical results for practical algorithms, respectively. 

We model the optimal caching problem as  an MAB problem \cite{MAB:P.Auer2002}. The MAB formulation can be used to model problems in which the system is partially known by the decision maker, and each action provides a different balance between maximising the instantaneous reward and acquiring new knowledge.  The original MAB problem considers a slot machine with several arms. At each time instant one arm is pulled, and the arm yields a random reward. The arms' rewards are independent and identically distributed (iid) random variables with unknown means. The expected values of the arms are estimated based on  past observations. The more times an arm is pulled the more reliable its estimate is, while the more times the arms with higher expected rewards are pulled the higher the expected accumulated reward is. Hence, there is a tradeoff between the exploration of new arms and the exploitation of known arms. 

If the arms' expected rewards were known, the optimal algorithm would pull, at each time slot, the arm with the highest expected reward. The  \emph{regret} of an algorithm is the  difference between its expected accumulated reward and that of the algorithm that always pulls the best arm. Hence, the  regret is a measure of the loss due to not knowing the reward profile of the arms. Literature on the MAB problem studies algorithms to decide which arm to  pull at each time instant in order to maximise the  accumulated expected reward, i.e., minimise the regret,  over time while balancing the exploration-exploitation tradeoff.

In \cite{MAB:T.Lai1985}, Lai and Robbins show that no algorithm can achieve an asymptotic regret smaller than $O(\log(t))$; that is, if the arms' rewards are not known the accumulated loss of the best algorithm grows at a logarithmic rate for large $t$. Notice that, although the accumulated loss is unbounded, since the logarithm grows slowly for large $t$, this bound suggest that the performance of the best algorithm can be very close to that of the optimal.  In \cite{MAB:P.Auer2002} one such algorithm, called the upper confidence bounds (UCB), is presented, and proven to achieve a regret on the order of $O(\log(t))$ uniformly over~$t$. An extension of the original MAB problem in which several arms can be pulled simultaneously is known as the combinatorial MAB (CMAB) \cite{MAB:W.Chen2013}. The combinatorial UCB (CUCB) algorithm proposed in \cite{MAB:W.Chen2013} is proven to achieve a regret behaviour of the order $O(\log(t))$ uniformly over time. In our system model replacing a file stored in the cache for a new one has a cost, which we model as an MAB problem with arm switching costs (MABSC). In \cite{MAB:R.Agrawal1990} an algorithm  that groups the arms' samples in order to minimise the number of arm switches is proven to achieve the asymptotic regret of order  $O(\log(t))$ for a special case of the combinatorial MABSC (CMABSC), in which a fixed number of arms is played at each~time.


\section{System Model} \label{CAsec:SM}

We consider  the content placement problem in an infostation, assuming that placing content into the infostation's cache memory has a cost, and the content popularity profile is not known in advance. The CC is in charge of deciding which are the best files to store in the cache. Users in the coverage area can offload some of their downlink traffic to the infostation by accessing the content stored in the cache memory. The infostation periodically broadcasts information about the cache content to its users; hence, the users readily know  the cache content. When a user wants to access a high data rate content, the request is directed to the infostation, if the content is in the cache; otherwise, it is downloaded directly from the cellular network. This process is carried out completely  transparent to the users; for example, through a smartphone application running in the background, that listens to the infostation broadcast signals and sends the user's request either to the infostation or to the cellular network, depending on the cache state. Note that the CC observes only the requests for files stored in the cache. 

The  infostation has a total cache memory of capacity $M$ units. We denote by $\mathcal{F}$ the set of all the files in the system, by $F=|\mathcal{F}|$ the total number of files, and by $S_f$ the size of the $f$th file in $\mathcal{F}$. Time is divided into periods, and we denote by $d^t_f$ the instantaneous demand for file $f$, that is, the number of requests for file $f$ in period $t$ normalised by $U$, where $U$ is the maximum number of users the infostation can serve at any given period.  The instantaneous demand, $d^t_f$, is an iid random variable with bounded support in $[0,1]$ and mean $\theta_f$.  We denote the content popularity profile by $\pmb{\Theta}=(\theta_1,\ldots,\theta_F)$. If a user requests a file $f$ that is stored in the cache, a ``hit'' is said to have occurred, and the file is downloaded directly from the infostation. We consider a reward of $S_f$ units when file $f$ is fetched from the infostation. This reward can be considered as a QoE gain for the user, or a bandwidth alleviation on the cellular system. At each period the CC updates the cache contents based on the demand history, where adding file $f$  has a cost of $S_f$ units. This cost corresponds to the bandwidth consumption on the infostation backhaul. 

The  aim of the CC is to optimize the cache content at each time period in order to maximise the traffic offloaded to the infostation, taking into account the cost associated to placing a file in the cache, and by simply observing the requests corresponding to the files in the cache over~time. 

A policy $\pi$ is an algorithm that chooses the cache content at each time period $t$, based on the whole history of the instantaneous demands and cached files. We denote the cache content in period $t$, chosen according to $\pi$, by  $\mathcal{M}^t_\pi$. We assume that $\mathcal{M}^0_\pi=\emptyset$, that is, the cache is initially empty. We denote  the instantaneous reward for file~$f$, stored in the cache, by~$r^t_f=U d_f^t S_f $. 

The expected  instantaneous total reward of policy $\pi$ in period $t$ is 
\begin{equation}\label{CAeq:rw} \eqnsize
  r_{\pmb{\Theta}}(\mathcal{M}^t_\pi)=\expected{\sum_{f\in \mathcal{M}^t_\pi }  U d_f^tS_f }{}=U\!\!\sum_{f\in \mathcal{M}^t_\pi} \!\! S_f\theta_f,
\end{equation}\normalsize
where the expectation is taken over the files' instantaneous demands. The cost associated with storing file $f$ into the cache is $S_f \cdot \mathbb{I}\{f\in \mathcal{M}^t_\pi ,f\notin \mathcal{M}^{t-1}_\pi\}$, where $\mathbb{I}\{a\}=1$  if $a$ is true, and $\mathbb{I}\{a\}=0$ otherwise.  The total cost of policy $\pi$ in period $t$ is
 \begin{equation}\eqnsize
  c(\mathcal{M}^t_\pi,\mathcal{M}^{t-1}_\pi)= \sum_{f\in \mathcal{M}^t_\pi} S_f \cdot  \mathbb{I}\{f\in \mathcal{M}^t_\pi ,f\notin \mathcal{M}^{t-1}_\pi\}.
\end{equation}\normalsize
We define the \emph{cache efficiency} as the total amount of traffic offloaded to the infostation minus the total cache replacement cost. The focus of this paper is to find a policy $\pi$ that maximises the cache efficiency over a time horizon $N$. This problem can be expressed as follows 
\begin{equation} 
\begin{aligned} \label{CAeq:opt_PG}\eqnsize
\max_{\pi}~ &  \sum_{t=1}^{N} \left [ r_{\pmb{\Theta}}(\mathcal{M}^t_\pi) - w\cdot c(\mathcal{M}^t_\pi,\mathcal{M}^{t-1}_\pi) \right ] \\
\text{s.t. } &
\sum_{f\in \mathcal{M}^t_\pi} S_f \leq M, ~~t=1,\ldots,N,
\end{aligned}\normalsize
\end{equation}
were $w$ is a weighting factor that arbitrates the relative cost of the backhaul and access bandwidth and can be chosen depending on the network state.  

If the  popularity profile, $\pmb{\Theta}$, is known, (\ref{CAeq:opt_PG}) is solved for the initial period and the cache content  is not changed in the following periods. In this case, if the time horizon, $N$, is large enough, the switching cost in the initial period can be ignored, and maximising \eqref{CAeq:opt_PG} is equivalent to maximising the expected immediate reward \eqref{CAeq:rw} under the cache capacity constraints, and it studied in \cite{Blasco2014}. This problem is called the single-period optimization (SPO) problem. In particular, this is a \emph{knapsack problem}, which is known to be NP-complete, and in general, can be solved using branching algorithms, such as branch and bound, with an exponential  worst case complexity. In our particular case the knapsack problem fulfils the so called \emph{regularity condition}, which implies that the solution of its linear program relaxation can be obtained by a \emph{greedy} algorithm~\cite{Dantzig1957}. The greedy algorithm starts with an empty cache memory, and adds files sequentially, starting from the files with higher popularity, $\theta_f$, until the cache is full. Note that all files in the cache, but the last cached file, are complete.  A $\left(1+O\left(\frac{1}{F}\right)\right)$-approximate solution to the knapsack problem is obtained by discarding this partially cached file from the cache~\cite{Korbut2010}. 

In Section~\ref{CAsec:CMAB} we assume the existence of an \mbox{($\pmb{\alpha},\pmb{\beta}$)-solver} for the SPO-problem. The  \mbox{($\pmb{\alpha},\pmb{\beta}$)-solver}, for $0\leq \alpha, \beta \leq 1$, is an algorithm which, for each popularity profile, outputs a set of contents. The expected reward of the algorithm output is,  with probability $\beta$, at least $\alpha$ times the optimal~reward.

Our main focus is on the more interesting case in which $\pmb{\Theta}$ is not known in advance, and has to be estimated. This problem is challenging since the instantaneous reward for files not cached in the infostation is not observed, and the CC can obtain information on the popularity of a specific content only by caching it, while caching new content has a cost. The CC wants to explore as many files as possible to discover the most popular ones, but it also wants to  exploit the limited storage capacity by caching the files that it believes to be the most popular, and keep the cache content static to minimise the cache cost. This is the well-known \emph{exploration vs. exploitation tradeoff}

\section{Learning the optimal cache content: regret bound }\label{CAsec:CMAB}

\subsection{CMABSC for optimal caching}
In problem (\ref{CAeq:opt_PG}) each file corresponds to one arm in the MAB problem, and a feasible cache content allocation corresponds to a feasible arm combination. At each period the CC decides the cache content according to policy $\pi$ (i.e., pulls the arms in $\mathcal{M}^t_\pi$), pays a cost $S_f$ for each new content $f$ added to the cache memory, and observes the instantaneous demands for the files in the cache, i.e., $r_f^t, \forall f \in \mathcal{M}^t_\pi$.  Since the instantaneous demand for each file is iid over time with an unknown mean value, and only the demands associated to $\mathcal{M}^t_\pi$ are observed, (\ref{CAeq:opt_PG}) is a CMABSC problem.

The  \emph{regret} of a policy $\pi$ is the  difference between its expected accumulated reward and that of the optimal policy, which knows the popularity profile, and caches the optimal content according to the ($\alpha,\beta$)-solver. We divide the regret into two parts, the \emph{sampling-regret} and the \emph{switching-regret}, which account for the loss due to not knowing the popularity profile, and the loss due to switching arms, respectively. We define the \emph{sampling-regret} of policy $\pi$ until period $t$ as
\begin{equation} \label{CAeq:sampling-regret}
 R^\pi_{Sa}(t)=t\alpha\beta r_{opt}-\expected{\sum_{i=1}^{t}r_{\pmb{\Theta}}(\mathcal{M}^i_\pi)}{},
\end{equation}
where $r_{opt}$ is the expected reward of caching the optimal set of files, that is, the optimal solution of the SPO-problem, and the expectation is taken over $\pi$ and the arms' rewards. Since with probability $\beta$ the \mbox{($\alpha,\beta$)-solver} finds a solution whose reward is at least $\alpha r_{opt}$, and the rewards are iid, the expected reward of the optimal policy at each period is at least $\alpha\beta r_{opt}$.   The \emph{switching-regret} of policy $\pi$ until period $t$  is  given by
\begin{equation}
\begin{aligned}
R^\pi_{Sw} (t)=\expected{\sum_{i=1}^t c(\mathcal{M}^i_\pi,\mathcal{M}^{i-1}_\pi)}{}-M.
\end{aligned}
\end{equation}
Notice that, since the cache is initially empty, the optimal policy incurs an initial switching cost of $M$ units. The  \emph{regret with switching cost}, $R^\pi(t)$, is the sum of the sampling and switching regrets: 
\begin{equation} \label{CAeq:regret}
\begin{aligned}
 R^\pi(t)=R^\pi_{Sa}(t) +w R^\pi_{Sw}(t).
 \end{aligned}
\end{equation}

The objective is to find a policy $\pi$ whose regret is small uniformly over $t$, i.e., grows sub-linearly with $t$, for all $t$.

\subsection{CUCB algorithm with switching cost } 
Classical algorithms for the MAB problem, such as UCB \cite{MAB:P.Auer2002} and CUCB \cite{MAB:W.Chen2013}, rely on the fact that an arm is considered \emph{well sampled} if it has been sampled/played more than a certain number of times, which depends on time $t$ and the arm's expected reward. In order to ensure that the arms are well sampled, the sample mean reward of each arm is perturbed with an additive positive term that increases the sample mean of the less often played arms. The arms with higher perturbed mean reward are played at each period.   Regret bounds for these algorithms do not depend on the times the arms are played. However, when switching costs are introduced, the time when an arm is played becomes important for the computation of the regret bound. We propose an algorithm, called the CUCB with switching cost (CUCBSC), that groups the samples of each arm into consecutive periods, such that the potential number of arm switches is small. The specific embodiment of CUCBSC is given in Algorithm \ref{alg:CUCBSC}. 

Time periods are divided into switching and non-switching periods. Arms are switched only in the switching periods, and in consecutive non-switching periods the same arms are played. The $b$th switching period occurs at time $t\!=\!n_b$, and the first switching period is  $n_1\!=\!F\!+\!1$. 
We define $\Delta(b)\! \triangleq \!n_{b+1}\!-\!n_{b}$. At period $t\!=\!n_b$ ($b$th switching period), some arms are switched and played, and these same arms are played until period $t\!=\!n_b\!+\!\Delta(b)-1$. Let $T_f$ denote the number of times arm $f$ has been played so far, $\pmb{\hat{\Theta}}=(\hat{\theta}_1,\ldots,\hat{\theta}_F)$ denote the sample mean estimate of $\pmb{\Theta}$, and  $\tilde{\theta}_f$ be the perturbed version of $\widehat{\theta}_f$.

 \begin{algorithm}[h!]
\small\caption{\mbox{CUCBSC}} \label{alg:CUCBSC}
\begin{algorithmic}
   \STATE \textbf{$\mathbf{1.}$ Initialize:}
   \STATE  cache all files at least once, observe the rewards, $r_f^t$, and update $\widehat{\theta}_f$ and $T_f$, $\forall f \in \mathcal{F}$.
   \STATE  set $b\leftarrow1$, and $t\leftarrow F+1$
   \STATE \textbf{$\mathbf{2.}$ Switching period $b$ (period $t=n_b$) :}
   \STATE  $\tilde{\theta}_f\leftarrow \widehat{\theta}_f+\sqrt{\frac{3 \log (t) }{2 T_f}}$, $\forall f \in \mathcal{F}$.
   \STATE  use $\tilde{\theta}_f,~\forall f,$ and the \mbox{($\alpha,\beta$)-solver}, to solve the SPO-problem for period $t$, obtain $\mathcal{M}^{t}$, and cache files in $\mathcal{M}^{t}$.
   \STATE \textbf{$\mathbf{3.}$  Non-switching periods:}
   \FOR{$\Delta (b)$ periods}          
           \FOR{ all $f \in \mathcal{M}^{t}$}
           \STATE observe reward, $r_f^t$
           \STATE set $\widehat{\theta}_f\leftarrow\frac{\widehat{\theta}_f\cdot T_f+\frac{r^t_f}{U\cdot S_f}}{T_f+1}$, and $T_f\leftarrow T_f+1$         
           \ENDFOR
           \STATE set $\mathcal{M}^{t+1}\leftarrow \mathcal{M}^{t}$, and   $t\leftarrow t+1$  
   \ENDFOR
   \STATE set $b \leftarrow b+1$, and go to Step $\mathbf{2.}$
  \end{algorithmic}
\end{algorithm}

Similar to the UCB and CUCB algorithms, the CUCBSC does not use the estimates $\widehat{\theta}_f$ to solve the SPO-problem, instead it uses the perturbed versions $\tilde{\theta}_f$. The perturbation consists of an additive positive term, whose square grows logarithmically with $t$, and the term itself decreases linearly with $T_f$. The perturbation promotes arms that are not played often by artificially increasing their expected reward estimates.   Notice that in step $2$, the best set of files are cached by using the \mbox{($\alpha,\beta$)-solver} and assuming $\tilde{\theta}_f$ is the true file popularity profile. Differently from the UCB and CUCB algorithms,  in  CUCBSC arms are switched only in a switching period, which reduces the switching-regret. Notice that step~$1$ in Algorithm \ref{alg:CUCBSC} can be avoided by using some prior popularity estimates, for example, obtained from the content provider.  

\subsection{CUCBSC regret bounds}
In this section we find non-trivial regret bounds for the CUCBSC algorithm. A sketch of the proof is provided in this section and the complete proof is relegated to Appendix \ref{CAapp:proof_regret}. 

First we introduce some definitions that will be useful for the proof. We denote by $\mathcal{G}$ the set of good arm combinations, i.e., $\mathcal{G}= \{ \mathcal{M}^t_\pi | r_{\pmb{\Theta}}(\mathcal{M}^t_\pi) \geq \alpha r_{opt} \}$, and by $\mathcal{B}$ the set of bad arm combinations, i.e.,   $\mathcal{B}= \{ \mathcal{M}^t_\pi | r_{\pmb{\Theta}}(\mathcal{M}^t_\pi) < \alpha r_{opt}\}$.  We define  $\Delta_u \triangleq \alpha\cdot r_{opt} - \min_{\mathcal{M}} \left \{ r_{\pmb{\Theta}} ({\mathcal{M}}) | {\mathcal{M}} \in \mathcal{B} \right \} $ and $\Delta_l \triangleq  \alpha\cdot r_{opt} - \max_{\mathcal{M}} \left \{ r_{\pmb{\Theta}} ({\mathcal{M}}) | {\mathcal{M}} \in \mathcal{B} \right \} $.  Note that there exists a linear function $g(\cdot)$, such that, $\left | r_{\pmb{\Theta}}({\mathcal{M}}^t_\pi)-r_{\pmb{\Theta}'}({\mathcal{M}}^t_\pi)\right |\leq g(\Lambda)$ if $\max_{1\leq i \leq F } |\theta_i-\theta_i'|\leq \Lambda$.

Each time a bad arm combination is played we say that a bad period has occurred. We denote by $N_{f,t}$ a counter that for each arm is updated only at bad periods. At each bad period $t$, if $f$ is the arm in $\mathcal{M}^t_\pi$ with the smallest $N_{f,t-1}$ value, we update $N_{f,t}=N_{f,t-1}+1$. In case of a draw, only one of the counters with the smallest value is chosen arbitrarily and incremented. With this update rule, $\sum_{f\in\mathcal{F}} N_{f,t}$ corresponds to the number of bad periods until time $t$, and we denote its expected value by $\overline{N}_t =  \expected{\sum_{f\in\mathcal{F}} N_{f,t}}{}$. Using similar techniques as in \cite{MAB:W.Chen2013} and \cite{MAB:J.Komiyama2013} we can obtain the following bound 
\begin{equation} \label{CAeq:EN_bound} \eqnsize
\overline{N}_t \leq \!(1-\beta) (t-F)+F \!\!\left ( K_1\!+\!\frac{6 \log t}{(g^{-1}(\Delta_{l}))^2}+\max_{1\leq j\leq b} \Delta (j)\! \right), 
\end{equation}\normalsize
where $b$ is the number of switching periods until time $t$, that is, $n_b \leq t < n_{b+1}$, and $K_1 = \sum_{j=1}^{\infty} 2 \cdot n_j^{-2} \Delta(j)$. The intuition behind (\ref{CAeq:EN_bound}) is that, apart from those due to the imperfect solver, the total number of bad periods grows with $\log t$.   Hence, for large $t$, the number of bad periods increases slowly. This guarantees higher rewards, but also that all the arms, including those in $\mathcal{B}$, are played with a certain non-zero probability.  Using  (\ref{CAeq:sampling-regret}) and (\ref{CAeq:EN_bound}) we bound the sampling-regret as follows:
\begin{equation} \eqnsize  \label{CAeq:sampling-regret-2}
R^\pi_{Sa}(t) \leq \left(K_1+\frac{6 \log t}{(g^{-1}(\Delta_{l}))^2}+\max_{1\leq j\leq b} \Delta (j) \right ) F  \Delta_{u}.
\end{equation}\normalsize
 We denote by $M_u$ the maximum cost of switching between two arm combinations, that is, $M_{u}= \displaystyle \max_{\mathcal{M}, \widehat{\mathcal{M}}} \{ c(\mathcal{M}, \widehat{\mathcal{M}}) \}$,  by $M_{l}$ the  maximum cost of switching between two good arm combinations, that is $M_{l}=  \displaystyle \max_{\mathcal{M}, \widehat{\mathcal{M}} \in \mathcal{G} }\{c(\mathcal{M}, \widehat{\mathcal{M}})\}$. We have $M_l \leq M_u$. To compute the switching-regret we count separately the number of switches between good arm combinations, and the rest of the arm switches.  The switching-regret is bounded by 
\begin{equation} \label{CAeq:Sw}
\begin{aligned}  \eqnsize
R^\pi_{Sw}(t) \leq \sum_{j=1}^b\frac{ \overline{N}_{n_{j+1}-1} -\overline{N}_{n_{j}-1}}{\Delta (j)} \cdot 2(M_u-M_{l}) +(b-1)\!\cdot \!M_{l} +F\!\cdot \! M_u,
\end{aligned}\normalsize
\end{equation}
where $n_b \leq t < n_{b+1}$. 

The growth rate of the sampling- and switching regrets is studied in Appendix~\ref{CAapp:proof_regret}. We show that both regrets grow logarithmically in $t$, and have an additional term that depends on $\Delta (b)$. 
We found that the switching-regret in (\ref{CAeq:Sw}) is bounded by a logarithmic function of $t$ plus the number of switching periods $b$, which is due to the fact that each switching period the \mbox{($\alpha,\beta$)-solver} outputs a bad arm combination with probability $1-\beta$.  If $\Delta (b)$ grows rapidly, then $b$ grows slowly, and it is possible to achieve a sub-linear switching-regret. On the other hand the sampling-regret in (\ref{CAeq:sampling-regret-2}) grows linearly  with  $\max_{1\leq j\leq b} \Delta (j)$ plus a logarithmic function of $t$. Intuitively, this implies that the more switching periods the better sampled the arms are; however, this implies a larger switching cost. To reduce the switching cost one can increase the number of non-switching periods (i.e., increase $\Delta(j)$), which in turn, implies a higher sampling regret. We conclude that there is a tradeoff between sampling and switching regrets, and that, if $\beta\neq 1$, in order to achieve a sub-linear regret, $\Delta(j)$ has to grow sub-linearly in $t$.  

\begin{thm} \label{CAthm:sqrt} The regret with switching cost of the CUCBSC algorithm  with $\Delta(j)=\left \lceil \gamma \sqrt{n_j}\right \rceil$ and $\frac{F^2+F-1}{\sqrt{F+1}}\geq \gamma \geq 2+\frac{1}{\sqrt{F+1}}$  is bounded by
 \begin{IEEEeqnarray}[\eqnsize]{rCL} 
\!\!R^{\pi}(t)\!\!&\leq&  \log t \!\left [\!\frac{6}{(g^{-1}(\Delta_{l}))^2}\!\! \left(\! w\! +\frac{\Delta_u}{2(M_u-M_l)}\right)\!\!+\!\!w\frac{\gamma}{2} \right]\!\cdot\!2\! F(M_u\!-\!M_l)  \nonumber\\
&+& \sqrt{t} \left[ wM_u+w(M_u-M_l)(1-2\beta) +F\Delta_u\gamma \right] + \left( \frac{\pi^2}{3}+4.12\gamma +1\right ) F  \Delta_{u}+wC,  \nonumber
\end{IEEEeqnarray}
where $C$ is a constant defined in Appendix~\ref{CAapp:proof_regret}.
\end{thm}
We remark that in the CMABSC problem we consider a general class of reward functions which may be computationally complex to solve, and assume that an \mbox{($\alpha,\beta$)-solver} with a randomised performance is available. This generalisation combined with the switching cost increases the order of the regret bound, that grows from a logarithmic order in \cite{MAB:W.Chen2013} to the order of $O(\sqrt{t})$, in Theorem~\ref{CAthm:sqrt}. 
To the best of our knowledge, there are no algorithms that have been proven to achieve better bounds, uniformly over time, for CMABSC or MAB problem with switching costs.

Now assume that the \mbox{($\alpha,\beta$)-solver} provides the unique optimal solution, i.e., $\alpha=\beta=1$.   
\begin{thm} \label{CAthm:L} If $\beta=\alpha=1$, the regret with switching cost of the CUCBSC algorithm with $\Delta(j)=L$ is bounded by
\begin{IEEEeqnarray}[\eqnsize]{rCL} \label{CAeq:regret2}
 R^\pi(t) &\leq &\frac{6F \log t}{(g^{-1}(\Delta_{l}))^2} \left(\Delta_{u}+w\frac{2M_u}{L} \right ) + F\Delta_{u}\!\left(\frac{\pi^2}{3}\!+\!L\!\right)\!\nonumber \\
 &&+\frac{\!wF\!\cdot\!2M_u}{L}\!\left(\frac{\pi^2}{3L} +\frac{3}{2L}  -1 +\frac{\log \left( 1+\frac{L-1}{F+1}\right)}{(g^{-1}(\Delta_{l}))^2} \right).
\end{IEEEeqnarray}
\end{thm}
Notice that if  $\Delta (j)=L$, the CUCBSC algorithm is the same as the CUCB algorithm \cite{MAB:W.Chen2013}, but the arms are switched every $L$ periods, instead of at every period. In particular, if the switching cost is removed, (i.e., $w=0$) and the arms are switched at each period $t$ (i.e., $L=1$) we get the same regret bound in (\ref{CAeq:regret2}) as in \cite{MAB:W.Chen2013}. Theorem~\ref{CAthm:L} extends the result in \cite{MAB:R.Agrawal1990} by achieving logarithmic regret uniformly over $t$.  While in the CUCBSC algorithm arms are switched only at the switching periods, in the MAB problem with lock-up periods  arms can be switched only at certain times~\cite{MAB:J.Komiyama2013}. Hence, the regret bound in (\ref{CAeq:regret2}) with $w=0$  extends the result in \cite{MAB:J.Komiyama2013} by considering the CUCB algorithm for the CMAB  problem with constant lock-up periods.

\section{Learning the optimal cache content: practical algorithms} \label{CAsec:mab_app}

Despite the theoretical bound on the regret growth, CUCBSC can take many iterations to learn the optimal cache content. 
In practice, simpler algorithms can achieve better performance \cite{V.Kuleshov2000}. One example is the \emph{\mbox{$\epsilon$-greedy}} algorithm, which caches at each iteration the best set of files according to the demand estimate $\hat{\mathbf{\Theta}}$ with probability $1-\epsilon$, and a random set of files with probability $\epsilon$. Due to the potentially large number of arm switches the  \mbox{$\epsilon$-greedy} algorithm induces a high switching cost;  and hence, we propose the \emph{\mbox{$(\epsilon,\Delta)$-greedy} algorithm}, which, every $\Delta$ iterations, caches the best set of files with probability $1-\epsilon$, and a random set with probability $\epsilon$. Despite the fact that caching a random set of files with probability $\epsilon$ incurs a linear regret, the \mbox{$\epsilon$-greedy} algorithm has been used in many practical applications. 

We propose yet another algorithm, based on CUCBSC, which we call the \emph{modified CUCBSC} (MCUCBSC) algorithm. The performance of the CUCBSC algorithm depends on the observations of the users' instantaneous demands. The more users are in the system, the more accurate the observations are, and the faster the algorithms can learn. The MCUBSC algorithm exploits this fact and that the popularity profile follows a Zipf-like distribution with parameter $\rho$. In MCUBSC the perturbation in step 2 of CUCBSC is modified as follows
\begin{equation}
     \tilde{\theta}_f\leftarrow\widehat{\theta}_f+\frac{1}{F^\rho} \sqrt{\frac{3 \log (\overline{U}t) }{2\overline{U} T_f}},        
\end{equation}
where $\overline{U}$ is the average number of users. The factor $\frac{1}{F^\rho}$ promotes exploitation when the Zipf distribution is skewed, that is, when $\rho$ is large and there are few popular files. The index $T_f$ is multiplied by $\overline{U}$, that is, exploitation is also promoted when $\overline{U}$ is large. This reflects the fact that, in each period, $\overline{U}$ independent realisations of the reward distribution are observed. Parameter $\rho$ can be empirically approximated as in \cite{caching:Breslau1999}.   

A well known algorithm for content caching in wired networks is the \emph{least recently used} (LRU) algorithm. Each time a file is requested, if it is not already in the cache, LRU discards the least recently used file in the cache and replaces it with the requested file. In our model, since demands are observed only for those files in the cache, LRU is not applicable directly, and we instead consider the \emph{$\Delta$-myopic} algorithm, which, every  $\Delta$ periods, caches the files that have been requested at least once  within the last $\Delta$ periods, and randomly fills the rest of the cache. Note that while the \mbox{$\epsilon$-greedy} and the MCUCBSC algorithms learn form all the past history, the $\Delta$-myopic learns only form the past $\Delta$ periods. Numerical comparison of these three algorithms is presented in the next section.

\section{Numerical Results}\label{CAsec:res}
In this section the performances of the MAB algorithms presented in Sections~\ref{CAsec:CMAB} and \ref{CAsec:mab_app}, namely CUCBSC, MCUCBSC, $\epsilon$-greedy, \mbox{\mbox{$(\Delta,\epsilon)$-greedy}}, and $\Delta$-myopic, are studied in an infostation terminal that provides high data rate service to its users.
The greedy approximation is used as the \mbox{($\mathbf{\alpha},\mathbf{\beta}$)-solver}. A number of numerical results involving different system parameters, such as the popularity profile ($\mathbf{\Theta}$), the average number of infostation users ($\overline{U}$), the cache memory size ($M$), and the total number of files in the system ($F$) are presented. 

To illustrate our results numerically we consider, unless otherwise stated, an infostation with a cache capacity of $M\!=\!512$ units, a total number of files $F\!=\!400$, and a maximum number of users $U=50$. We assume that the set of file sizes is $\{2^i\}$, for $i=\{0,1,\ldots,7\}$, and that there are $50$ files of each size. We assume that each period a random number of users, uniformly distributed in $[0,U]$,  within the infostation coverage area request a file following a Zipf-like distribution with skewness parameter $\rho=0.56$ (same as in \cite{caching:Golrezaei2012a} and \cite{caching:Breslau1999}).  Notice that, similar to \cite{caching:Poularakis2013}, the cache memory can only store approximately $4\%$ of the total available content at any given time. In the rest of the paper, if the size of the cache is given in percentage, it is referred to the percentage of the total content size that can be stored in the cache memory at any given time. We consider that the access and backhaul costs are balanced, and assume $w=1$ in \eqref{CAeq:opt_PG}. We study the CUCB algorithm as well as the two modified versions with reduced switching cost, namely the CUCBSC algorithm with $\Delta(b)=10$ and $\Delta(b)=\lceil 2 \sqrt{n_b} \rceil$, the $\epsilon$-greedy algorithm with $\epsilon=0.1$, and the  $(\Delta,\epsilon)$-greedy algorithm with $(\Delta,\epsilon)=(10,0.1)$, the  MCUCBSC algorithms with $\Delta(b)=10$ and $\Delta(b)=\lceil 2 \sqrt{n_b} \rceil$, and the $\Delta$-myopic algorithm with $\Delta=10$.  

\begin{figure}[!ht]
  \begin{center}
        \subfigure[]{%
           \label{CAfig:RSC}
\includegraphics[width=0.36\textwidth]{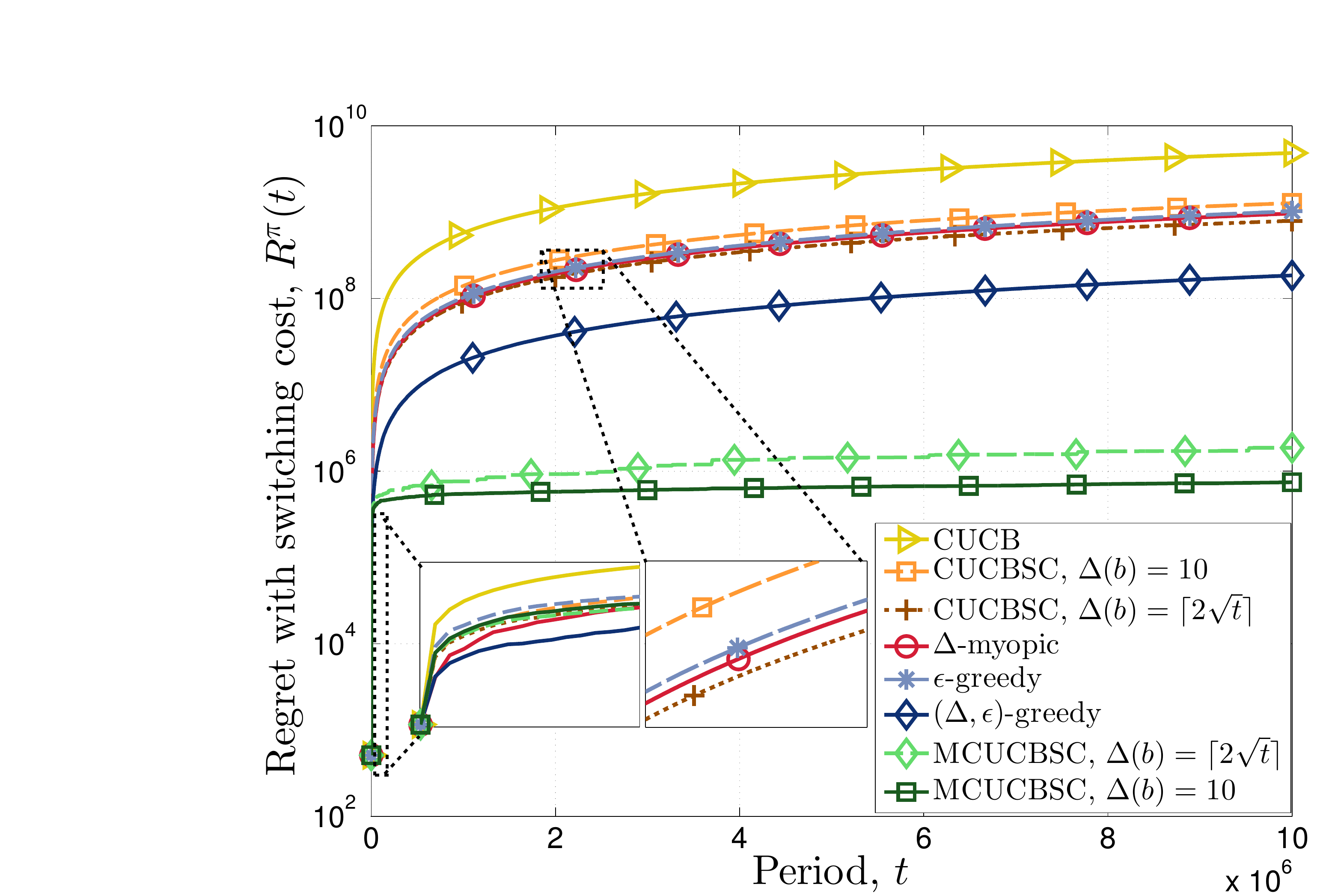}
        }%
        \\
      \subfigure[]{%
           \label{CAfig:SWR}
\includegraphics[width=0.36\textwidth]{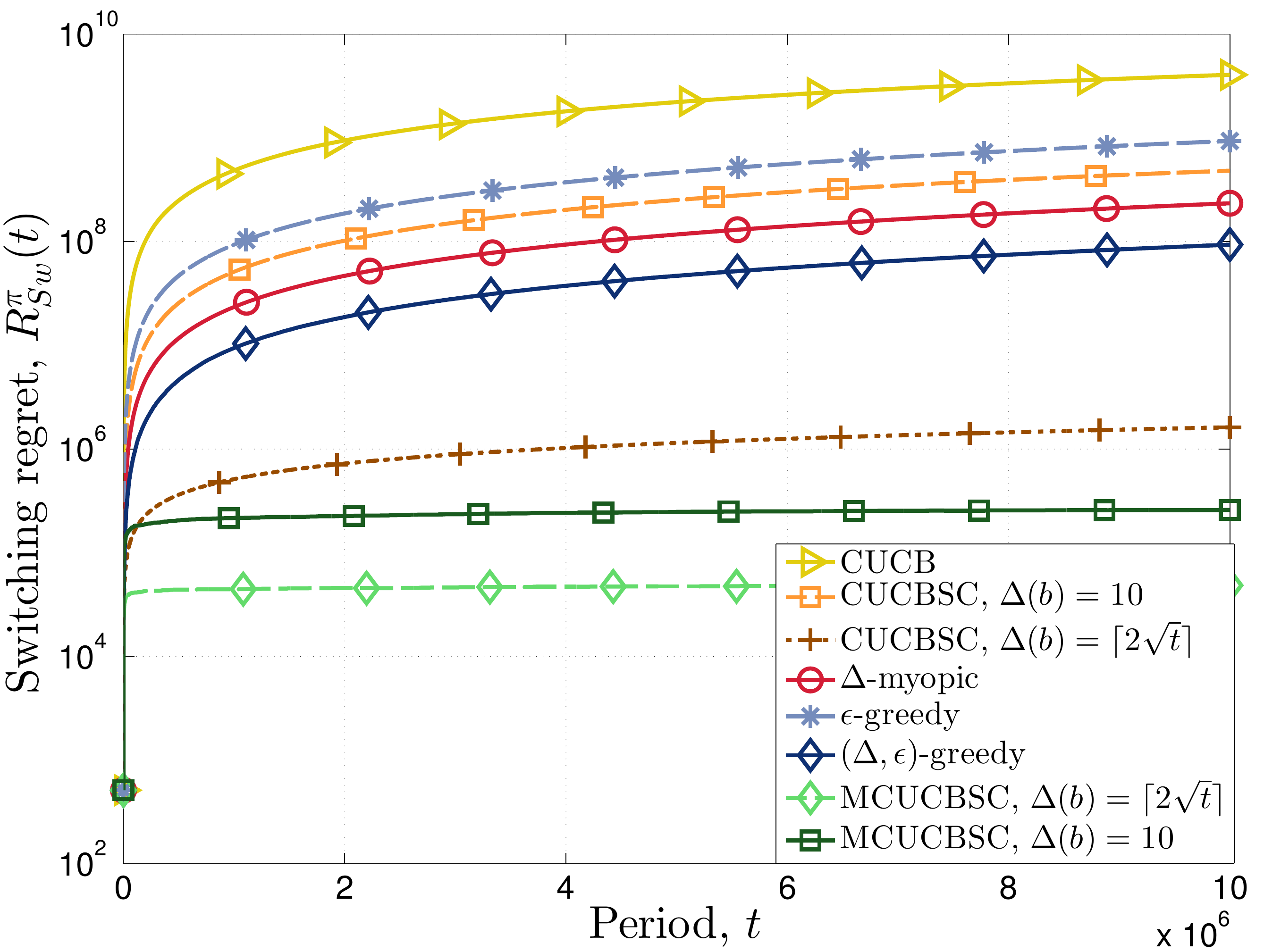}
        }%
        \subfigure[]{%
           \label{CAfig:SAR}
\includegraphics[width=0.35\textwidth]{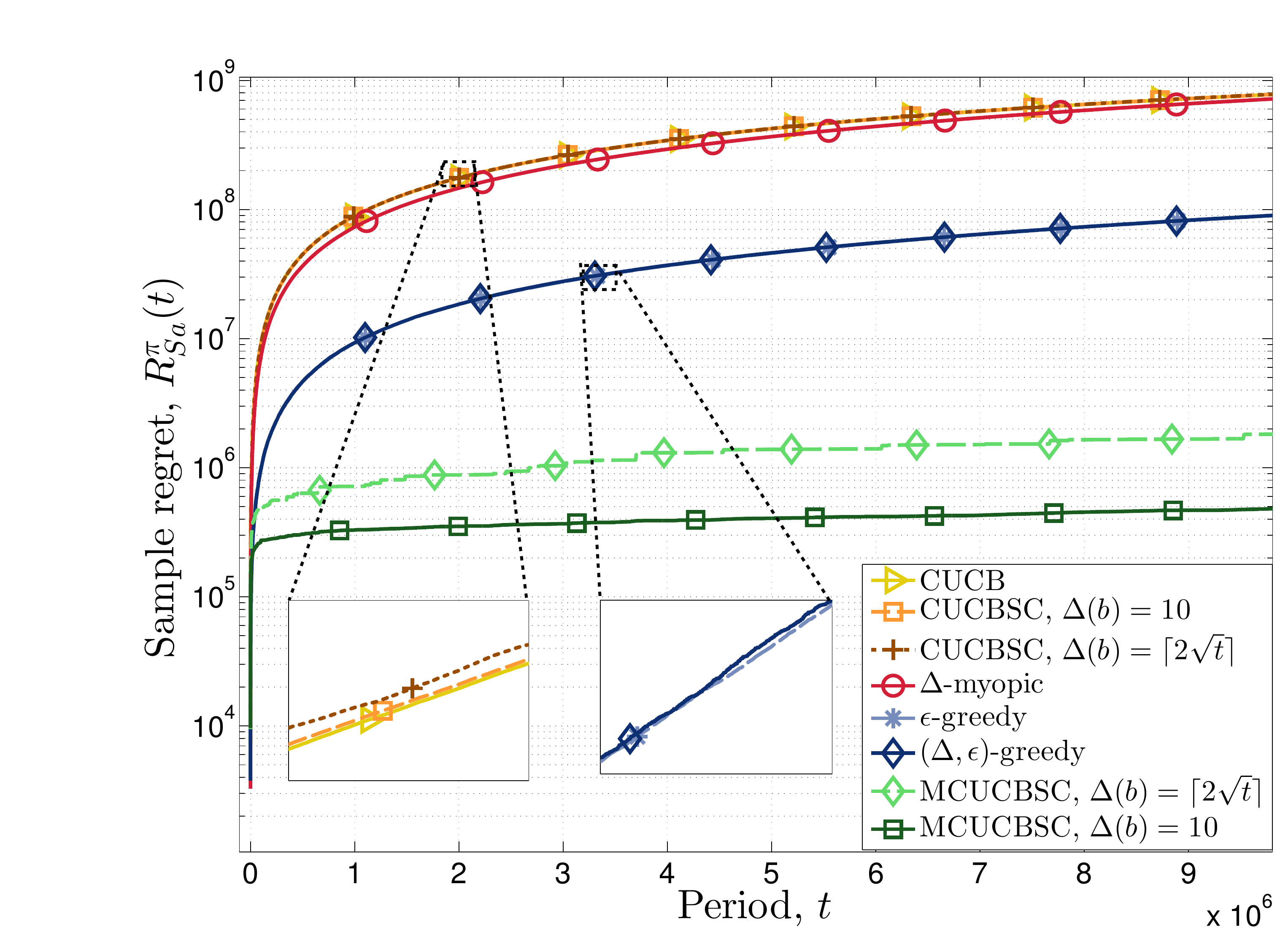}
       }
    \end{center} \vspace{-0.6cm}
    \caption{Regret with switching cost, and the switching and sampling regrets separately for the studied MAB algorithms.} \label{CAfig:Regret} \vspace{-0.6cm}
\end{figure}

Time evolutions of the regrets for these MAB algorithms are plotted in Figure~\ref{CAfig:Regret}.  In particular, in Figure~\ref{CAfig:RSC} we observe that, besides the lack of theoretical convergence results, the MCUCBSC algorithms have smaller regret than the CUCBSC algorithms. The growth of the regret is very small for the MCUCBSC algorithm with $\Delta(b)=10$, and it is practically steady after few periods.  

The results in Section \ref{CAsec:CMAB} indicate that there is a tradeoff between sampling and switching regrets. Figures~\ref{CAfig:SWR} and \ref{CAfig:SAR} show the switching and sampling regret, respectively, over time for the MAB algorithms.  We observe that the MAB algorithms that have smaller switching regret have larger sampling regret as well.  In Figure~\ref{CAfig:RSC}, we observe that CUCBSC and greedy algorithms that have less switching periods (i.e.,  $(\Delta,\epsilon)$-greedy and CUCBSC with $\Delta(b)=\lceil 2 \sqrt{n_b} \rceil$)  perform better than their counterparts (i.e., $\epsilon$-greedy, CUCBSC with $\Delta(b)=10$ and CUCB). Hence, for those algorithms, the loss due to not switching arms at each period is smaller when compared to the reduction in the switching regret. The opposite holds for the MCUCBSC algorithm, with $\Delta=10$, which has smaller regret than the MCUCBSC with $\Delta(b)=\lceil 2 \sqrt{n_b} \rceil$.

%
%

From this point onwards we study the performance of the MCUCBSC with $\Delta=10$, the \mbox{$(\Delta,\epsilon)$-greedy} algorithm  with $(\Delta,\epsilon)=(10,0.1)$,  and the $\Delta$-myopic algorithm with $\Delta=10$.  In addition to the MAB algorithms, we consider an informed upper bound (IUB) which assumes that the popularity profile is known in advance, and decides the cache content using the \mbox{$(\alpha,\beta)$-solver}.  We study the performance of the MAB algorithms as well as the IUB for a time horizon of $N=5\cdot 10^4$ periods, and average the results over $500$ experiments. We normalise the cache efficiency (\ref{CAeq:opt_PG}) by the total amount of data requested by the users until time horizon $N$, that is, data requested from the infostation as well as the cellular network. Hence the cache efficiency is measured as the percentage of net traffic that is offloaded to the infostation compared to the total traffic.

\begin{figure}[!ht]
     \begin{center}
       \subfigure[]{%
           \label{CAfig:gamma}
\includegraphics[width=0.36\textwidth]{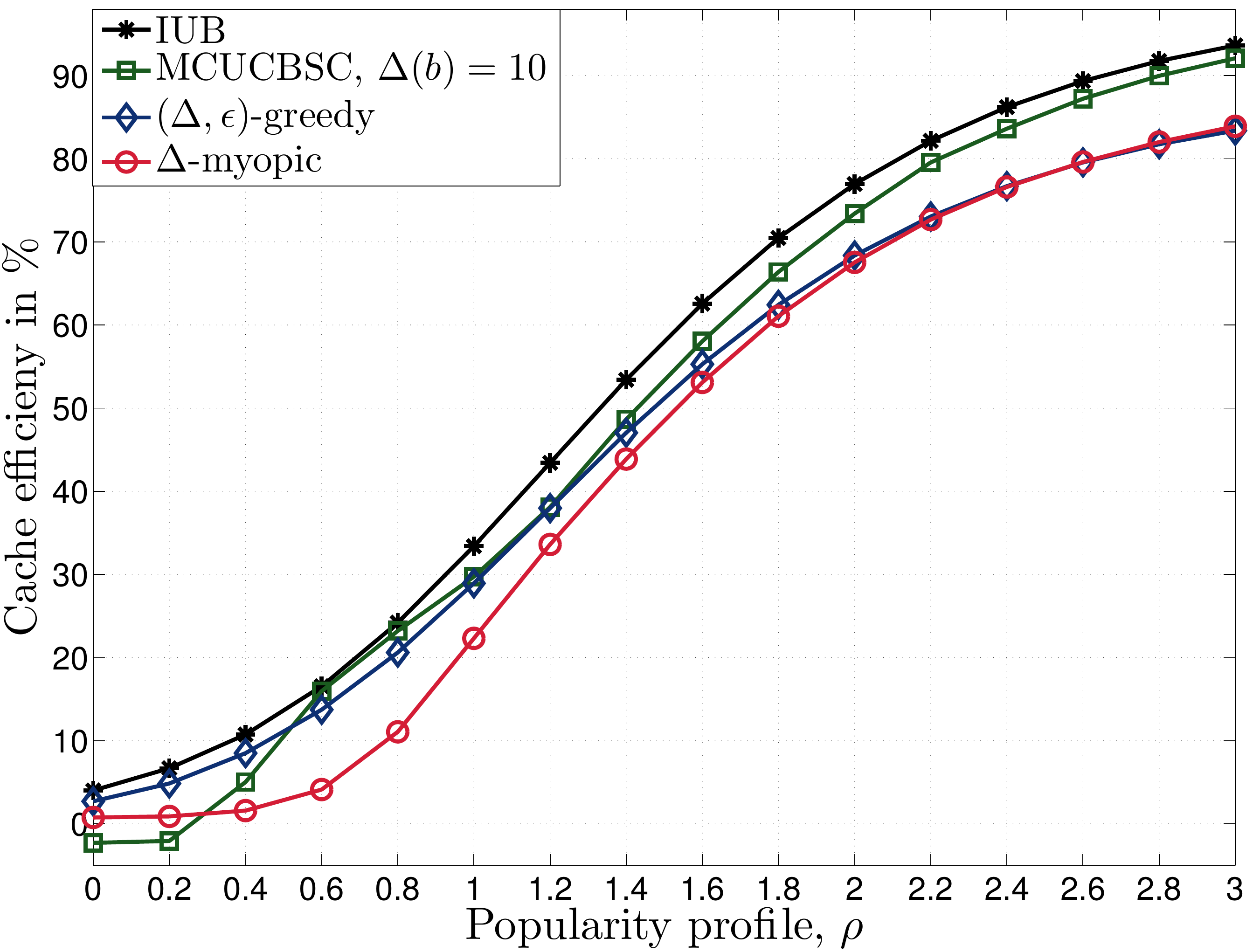}
        }%
        \subfigure[]{%
           \label{CAfig:cache}
\includegraphics[width=0.36\textwidth]{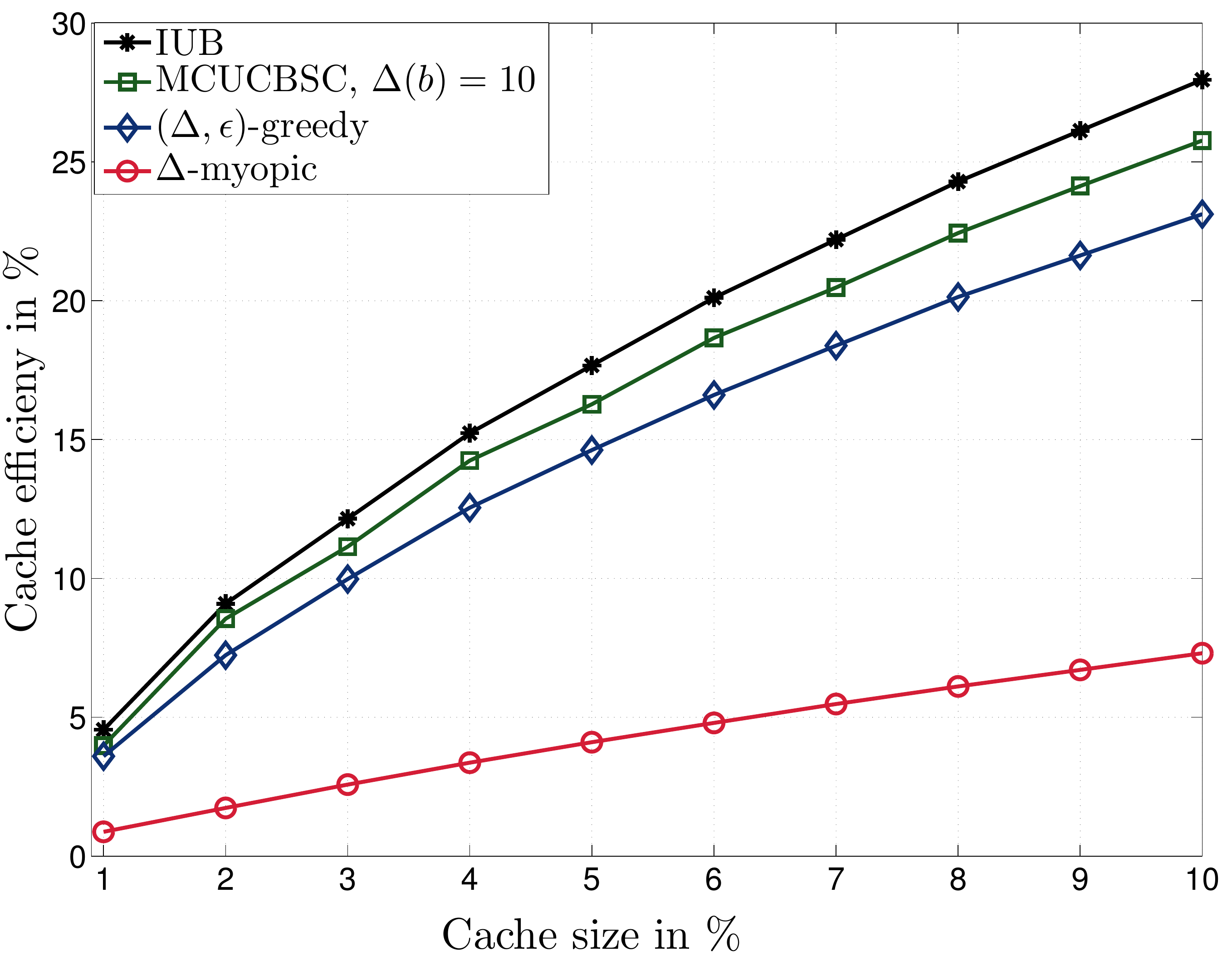}
       }
    \end{center} \vspace{-0.6cm}
    \caption{Cache efficiency for the MAB and IUB algorithms with respect to the Zipf distribution parameter ($\rho$), and the cache capacity ($M$).  }  \vspace{-0.6cm}
\end{figure}

Figure~\ref{CAfig:gamma} shows the effect of the popularity profile on the cache efficiency. Clearly, when the popularity profile is uniform, that is, when $\rho$ is small, all algorithms have low cache efficiency. In particular the IUB has a cache efficiency close to $4\%$, which is the relative size of the cache memory. This is due to the fact that, if the demand is uniform the composition of the cache content is irrelevant.  Due to the cache replacement cost the proposed algorithms have a lower cache efficiency compared to IUB. In particular, the MCUCBSC has a negative cache efficiency, that is, the cost of caching files is superior to the cost of serving user's requests from the cellular network directly. As the popularity profile becomes more skewed, IUB upper bound, and the cache efficiency of the proposed algorithms increase until they reach close to $100\%$ efficiency. Notice the  gap between \mbox{$(\Delta,\epsilon)$-greedy} and MCUCBSC, which is due to the constant exploration term, $\epsilon$.  The $\Delta$-myopic algorithm follows a similar trend, albeit with a lower cache efficiency when the popularity profile is more uniform.

The cache efficiency with respect to the cache size, measured in percentage of the total content size, is studied in Figure~\ref{CAfig:cache} for $\rho=0.56$. We note that as the cache size grows the cache efficiency increases as well, exhibiting a behaviour slightly below linear.  In particular, the IUB upper bound has a cache efficiency of $5\%$ for a cache size of $1\%$, that is, five times the cache size, while for a cache size of $10\%$ the cache efficiency ramps up to $27\%$, that is, only $2.7$ times the cache size.  A similar behaviour is observed for other algorithms. This is due to the Zipf-like popularity profile, in which file popularity decays rapidly, and the fact that caching popular files has higher cache efficiency than caching less popular files. Hence, for small cache sizes,  
only popular files are cached which has a high cache efficiency. In the other hand, for large caches sizes, less popular files are cached as well, which has lower cache efficiency. 

\begin{figure}[!ht]
     \begin{center}
       \subfigure[]{%
           \label{CAfig:users}
\includegraphics[width=0.36\textwidth]{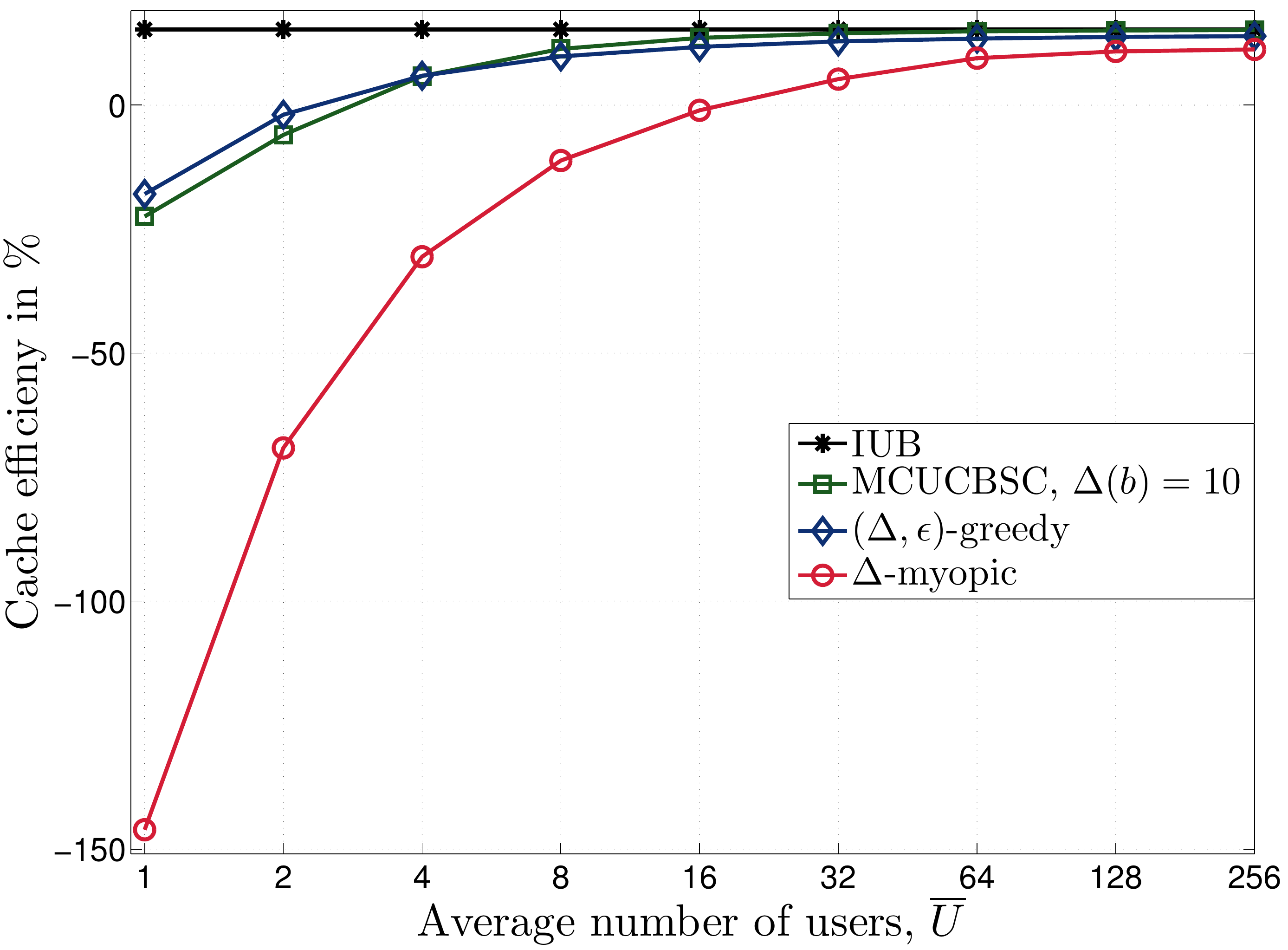}
        }%
        \subfigure[]{%
           \label{CAfig:file}
\includegraphics[width=0.36\textwidth]{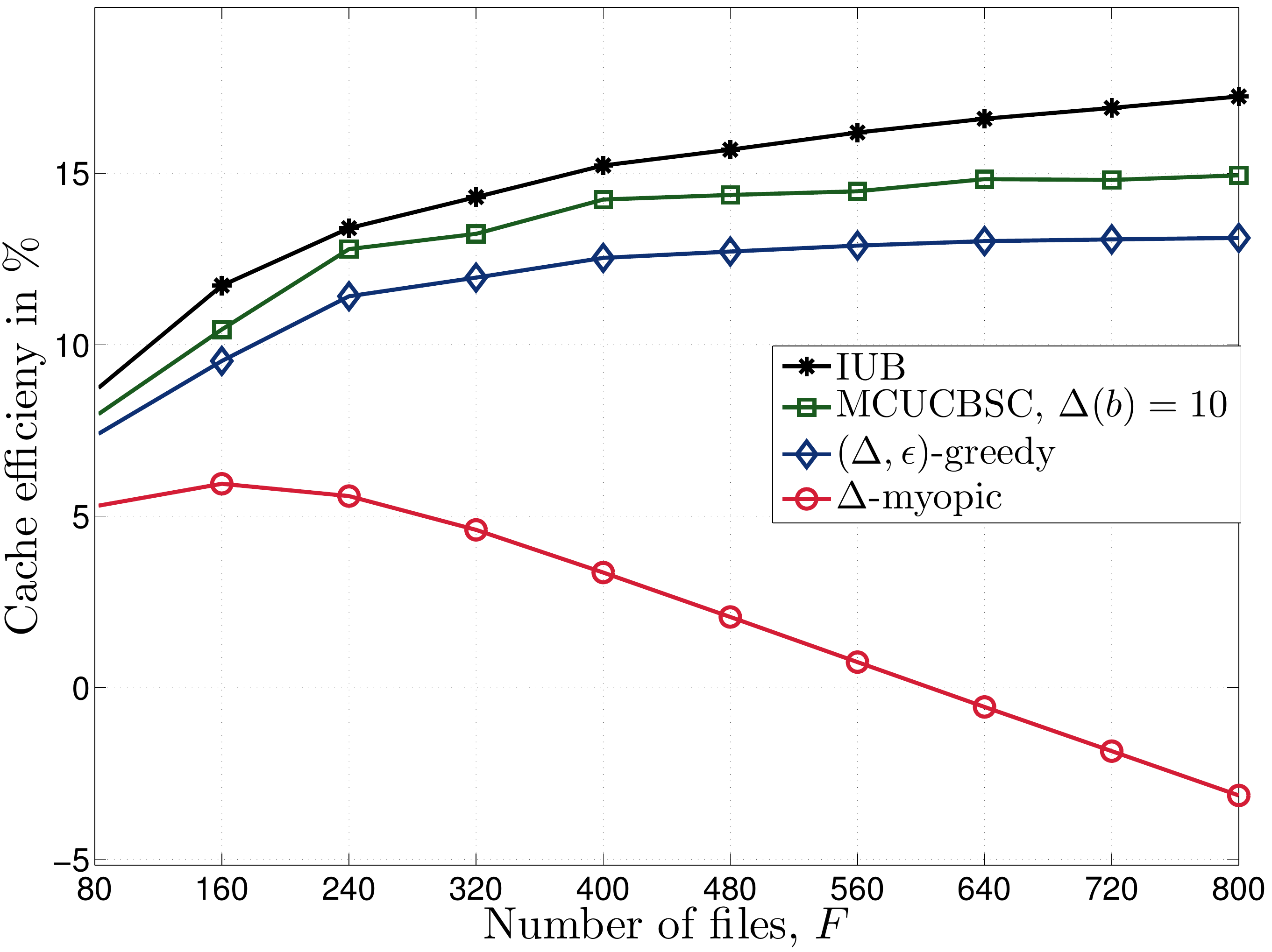}
       }
    \end{center} \vspace{-0.6cm}
    \caption{Cache efficiency for the MAB and IUB algorithms with respect to the average number of users ($\overline{U}$) and the file set size ($F$), for fixed cache size of $4\%$. }  \vspace{-0.6cm}
\end{figure}

The performances of the MCUCBSC, \mbox{$(\Delta,\epsilon)$-greedy}, and $\Delta$-myopic algorithms depend on the observations of the users' instantaneous demands. If the number of infostation users is low, the observations become less accurate, and the algorithms learn more slowly.  Figure~\ref{CAfig:users} depicts the performance of the proposed algorithms for different number of users. When the average number of users is low, i.e., $\overline{U}=\{1,2\}$, all the algorithms have negative cache efficiency. This is because the switching cost is high compared to the user data traffic for small $\overline{U}$.  The cache efficiency of the MCUCBSC and \mbox{$(\Delta,\epsilon)$-greedy} algorithms are negative for $\overline{U}\leq 4$, and close to the IUB upper bound  for $\overline{U}\geq 16$. This confirms that caching content in the network edges is more efficient in population dense areas. On the other hand the $\Delta$-myopic policy has negative cache efficiency for small values of $\overline{U}$, and  positive for  $\overline{U}\geq 16$.  This is due to the fact that, when $\overline{U}$ is small,  there are few  requests per period, and hence, the $\Delta$-myopic replaces the files often, incurring a large switching cost.  Note that the total traffic increases linearly with $\overline{U}$, and so does the traffic offloaded to the infostation. 

Finally, in Figure~\ref{CAfig:file} we study the effect of the number of files, $F$. We impose that, independent of $F$, the cache size can always hold approximately $4\%$ of the total available content. The popularity profile is more skewed and  has a wider peak for large and small $F$, respectively. Since the cache memory can store only $4\%$ of the files, when $F$ is small there are popular files that do not fit into the cache. The performance of the MCUCBSC and \mbox{$(\Delta,\epsilon)$-greedy} algorithms drop approximately by $5\%$  when $F$ is small, and grow steadily with $F$. The cache efficiency of the $\Delta$-myopic algorithm, contrary to that of the others, decreases with $F$. This is due to the fact that while both $F$ and the cache size increase, the average number of users' requests remains constant, hence, less ``hits'' occur and more files are replaced at each period,  incurring a larger switching cost for the $\Delta$-myopic algorithm.

 \section{Conclusions}\label{CAsec:con}
We have introduced the novel concept of \emph{content-level selective offloading} through the introduction of an \emph{infostation} terminal that stores high data rate content in its cache, and provides this content to its users directly, reducing the latency and the pressure on the congested backhaul links.  We have studied the optimal content caching problem when the file popularity profile is unknown and storing new content in the cache has a cost. The CC optimises the cache contents based on the demand history for the cached files in order to maximise the cache efficiency. We have modeled the problem as a combinatorial MAB problem with switching costs. To solve this problem, we have proposed the CUCBSC algorithm, and shown non-trivial regret bounds that hold uniformly over $t$. We have also proposed the MCUCBSC algorithm, which adapts to the special conditions of the optimal cache content problem by taking into account the file popularity profile and the number of users in the system; and the $(\Delta,\epsilon)$-greedy  algorithm, a version of the well known $\epsilon$-greedy algorithm that takes into account the switching cost. Our numerical results have shown that caching content at the network edge, and in particular in busy cellular systems, with large number of users, can bring large benefits and reduce the traffic in the backhaul network significantly.  We have observed that the cache efficiency increases with the content popularity skewness and the cache capacity.

\appendices
\section{Proofs of Theorems \ref{CAthm:sqrt} and \ref{CAthm:L}} \label{CAapp:proof_regret}
We introduce the Chernoff-Hoeffding's inequality, which will be used in the proof. 
\begin{thm} [Chernoff-Hoeffding's Inequality \cite{Hoeffding1963}] \label{CAthm:chernoff-Hoefding}
 Let $\mu$ be  an iid random variable with bounded support in $[0,1]$ and mean $\bar{\mu}$, and let  $\hat{\mu}_n$ be the mean of $n$ realizations of $\mu$. Then for any $a\geq0$, we have  $P\{|\hat{\mu}_n -\mu | \geq \sfrac{a}{n} \}\leq 2\cdot e^{\frac{-2 a^2}{n}}$.
\end{thm}
In the rest of the proof we ignore the under script $\pi$ in $\mathcal{M}^t_\pi$, since when we consider any policy other than $\pi$ it is clearly stated. 

\subsection{Sampling-Regret}  \label{CAapp:proof_sampling}
\subsubsection{Bound on the expected number of bad periods}
Consider a period $t$, for which $n_b \leq t < n_{b+1}$ for some $b$. Define $l_t=\frac{6\log t}{(g^{-1}(\Delta_{l}))^2}$. In addition to the counter $N_{f,t}$, which is updated at each bad period, we define the counter $T_{f,t}$, updated each time the arm $f$ is played $T_{f,t} \geq N_{f,t}$.

We define $\Delta^{b,t} (j)$, for $j \leq b$ and $n_b \leq t < n_{b+1}$, as follows  $\Delta^{b,t} (j)=\Delta(j)$ if $j<b$ and $\Delta^{b,t} (j)=t-n_b+1$  if $j=b$. We denote the event that the ($\alpha,\beta$)-solver outputs an arm combination that is not $\alpha$ times the optimal in period $t=n_b$, by $\chi_b$.  Then the  number of bad periods up to time $t$ is bounded by
\begin{subequations} \label{CAeq:A:N_bound}
\begin{IEEEeqnarray}[\small]{rcl} 
  \sum_{\mathclap{f=1}}^F N_{f,t} &\leq&  \sum_{\mathclap{n=F+1}}^{t} \mathbb{I} \{{\mathcal{M}}^{n} \in\mathcal{B} \} +F  \label{CAseq:BP1}\\
 &=& \sum_{\mathclap{j=1}}^{b}  \mathbb{I} \{{\mathcal{M}}^{n_j} \in\mathcal{B} \}\Delta^{b,t}(j) +F \label{CAseq:BP2}\\
 &=& \sum_{\mathclap{j=1}}^{b}\sum_{\mathclap{f=1}}^F \mathbb{I} \{{\mathcal{M}}^{n_j}\in \mathcal{B}, N_{f,n_{j}}>N_{f,n_{j}-1}\} \Delta^{b,t}(j) + F \label{CAseq:BP3}\\
 &\leq& \sum_{\mathclap{j=1}}^{b}\sum_{\mathclap{f=1}}^F \mathbb{I} \{{\mathcal{M}}^{n_j}\in \mathcal{B}, N_{f,n_{j}}>N_{f,n_{j}-1},N_{f,{n_{j}-1}}\geq l_{t}  \} \Delta^{b,t}(j) \nonumber\\
 & &+ F \left ( 1+l_t+\max_{1\leq j\leq b}\Delta^{b,t}(j) -1\right )\label{CAseq:BP4}\\
 &=& \sum_{\mathclap{j=1}}^{b}  \mathbb{I} \{{\mathcal{M}}^{n_j}\in \mathcal{B}, N_{f,{n_{j}-1}}\geq l_{t}, \forall f \in {\mathcal{M}}^{n_j}  \} \Delta^{b,t}(j)   + F \left (l_t+\max_{1\leq j\leq b}\Delta^{b,t}(j) \right ) \label{CAseq:BP5}\\
 &\leq& \sum_{\mathclap{j=1}}^{b}  \mathbb{I} \{{\mathcal{M}}^{n_j}\in \mathcal{B}, N_{f,{n_{j}-1}}\geq l_{n_j}, \forall f \in {\mathcal{M}}^{n_j}  \} \Delta^{b,t}(j)  + F \left (l_t+\max_{1\leq j\leq b}\Delta^{b,t}(j) \right ) \label{CAseq:BP6}\\
  &\leq& \sum_{\mathclap{j=1}}^{b} \left ( \mathbb{I} \{ \chi_j \} +  \mathbb{I} \{ \neg \chi_j, {\mathcal{M}}^{n_j}\in \mathcal{B}, N_{f,{n_{j}-1}}\geq l_{n_j}, \forall f \in {\mathcal{M}}^{n_j}  \} \right ) \Delta^{b,t}(j) \nonumber\\
  & &+ F \left (l_t+\max_{1\leq j\leq b}\Delta^{b,t}(j) \right ) \label{CAseq:BP7} \\
  &\leq& \sum_{\mathclap{j=1}}^{b} \mathbb{I}\{ \chi_j \}\Delta^{b,t}(j) + \sum_{\mathclap{j=1}}^{b}\mathbb{I} \{ \neg \chi_j, {\mathcal{M}}^{n_j}\in \mathcal{B}, T_{f,{n_{j}-1}}\geq l_{n_j}, \forall f \in {\mathcal{M}}^{n_j}  \}  \Delta(j)\nonumber \\
  & &+ F \left (l_t+\max_{1\leq j\leq b}\Delta(j) \right ), \label{CAseq:BP8}
\end{IEEEeqnarray}
\end{subequations}
where \eqref{CAseq:BP1} follows by assuming that no good arm combination is played  in the initalization of the CUCBSC algorithm; \eqref{CAseq:BP2} from the fact that arms are switched only in switching periods;  \eqref{CAseq:BP3} follows since we use that $N_{f,n_j}$ is updated only once each period;  the counters that are larger than $l_t$ are summed in the first line of \eqref{CAseq:BP4}, while the second line is an upper bound on the counters that are smaller than $l_t$; in \eqref{CAseq:BP5} we use the fact that only the smaller counter among the played arms  is updated at each bad period,  in \eqref{CAseq:BP6} we use the fact that $l_t$ is monotonically increasing in $t$, and that $t\geq n_j$; in the left-hand side of the first line in \eqref{CAseq:BP7} we count the bad periods due to errors in the \mbox{($\pmb{\alpha},\pmb{\beta}$)-solver}, and in \eqref{CAseq:BP8} we use that $T_{f,t} \geq N_{f,t}$ and $\Delta(j) \geq \Delta^{b,t}(j)$. 

To bound the expected number of bad periods, $\overline{N}_t$, we need to compute the probability of $\{ \neg \chi_j, {\mathcal{M}}^{n_j}\in \mathcal{B}, \forall f \in {\mathcal{M}}^{n_j},T_{f,{n_{j}-1}}\geq l_{n_j}  \}$ being true. This is, the probability of the event that all the arms in $\mathcal{M}^{n_j}$ have been sampled/played more than $l_{n_j}$ times, and the $(\alpha,\beta)$-solver has not failed, the switching period $n_j$  is a bad period. To obtain this probability similar arguments as in \cite{MAB:W.Chen2013}, which are based on Theorem \ref{CAthm:chernoff-Hoefding}, can be used. Let $a= \sqrt{\frac{3 \log n_j}{2 T_{f,n_{j}-1} }} \cdot T_{f,n_{j}-1} $ in Theorem~\ref{CAthm:chernoff-Hoefding}. We obtain $P\{ \neg \chi_j, \mathcal{M}^{n_j}\in \mathcal{B}, \forall f \in \mathcal{M}^{n_j}, T_{f,n_{j}-1}>l_{n_j}  \}\leq 2\cdot F \cdot n_j^{-2}$. Notice that the probability vanishes proportionally to $\frac{1}{{n_j}^2}$. This means that if all the arms are sampled often enough (more than $l_{n_j}$) the probability of a having a bad period vanishes for large $t$. Summing up these probabilities until switching period $b$ we get
\begin{subequations}\label{CAeq:A:k1}
\begin{IEEEeqnarray}[\eqnsize]{L} 
  \expected{\sum_{j=1}^{b}\mathbb{I} \{ \neg \chi_j, {\mathcal{M}}^{n_j}\in \mathcal{B}, \forall f \in {\mathcal{M}}^{n_j}, T_{f,{n_{j}-1}}>l_{n_j}  \}  \Delta(j)}{} \\
  \leq   F \sum_{j=1}^{b} 2 \cdot n_j^{-2} \Delta(j) \\
  \leq   F \sum_{j=1}^{\infty} 2 \cdot n_j^{-2} \Delta(j) =  F  \cdot K_1,
\end{IEEEeqnarray}
\end{subequations}
we have defined $K_1 \triangleq 2 \sum_{j=1}^{\infty} \frac{\Delta(j)}{n_j^2}$. 

To prove \eqref{CAeq:EN_bound}, that is,  an upper bound on the expected number of bad periods until period $t$, we use (\ref{CAeq:A:N_bound}), (\ref{CAeq:A:k1}), and the fact that $\expected{~\mathbb{I}\{ \chi_j \}}{}=1-\beta$
\begin{subequations}\label{CAeq:A:EN_bound}
\begin{IEEEeqnarray}[\eqnsize]{rCL} 
   \overline{N}_t 
   &\leq&  \expectedBig{\sum_{j=1}^{b} \mathbb{I}\{ \chi_j \}\Delta^{b,t}(j) + \sum_{j=1}^{b} \mathbb{I} \{ \neg \chi_j, {\mathcal{M}}^{n_j}\in \mathcal{B}, \forall f \in {\mathcal{M}}^{n_j}, T_{f,n_{j}-1}>l_{n_j}  \} \Delta(j)}{}  \nonumber \\   
   &&+ \expected{F \left ( l_t+\max_{1\leq j\leq b} \Delta(j) \right ) }{} \\
   &\leq& (1-\beta) (t-F)+F \left ( K_1+l_t+\max_{1\leq j\leq b} \Delta(j) \right ).
\end{IEEEeqnarray}
\end{subequations}
\subsubsection{Sampling-regret} 
To prove \eqref{CAeq:sampling-regret-2}, that is, the sampling regret, we plug (\ref{CAeq:A:EN_bound}) into \eqref{CAeq:sampling-regret}
\begin{subequations}\label{CAeq:A:sampling-regret}
\begin{IEEEeqnarray}[\eqnsize]{rCL} 
R^\pi_{Sa}(t)&= &t\cdot\alpha\cdot\beta\cdot r_{opt}-\sum_{i=1}^{t}r_{\pmb{\Theta}}({\mathcal{M}}^i_\pi)\\
&\leq & t\alpha\beta r_{opt}- \left ( t\alpha r_{opt}- \overline{N}_t\Delta_{u} \right )\\
& \leq &\left(K_1+l_t+\max_{1\leq j\leq b} \Delta(j) \right ) F  \Delta_{u}.  
\end{IEEEeqnarray}
\end{subequations}
 
\subsection{Switching-regret} \label{CAapp:proof_switching} 
 Now, we bound the cost related to switching arms. Notice that the cost of switching occurs only in the switching periods $t=n_b$, and that the switching cost remains constant until the next switching period $t=n_{b+1}$.  The expected switching cost until period $t$, $R^\pi_{Sw}(t)$, with $n_b \leq t < n_{b+1}$, is bounded by 
\begin{subequations}\label{CAeq:A:Sw}
\begin{IEEEeqnarray}[\small]{rCL} 
R&^\pi_{Sw}&(t) 
=\expected{\sum_{i=1}^t \sum_{f=1}^F S_f \cdot \mathbb{I} (\mathcal{M}^i,\mathcal{M}^{i-1})}{}  -M\label{CAseq:SC2}\\
&=& \expected{\sum_{j=1}^b \sum_{f=1}^F S_f \cdot \mathbb{I} \{f \notin {\mathcal{M}}^{n_{j-1}}, f \in {\mathcal{M}}^{n_j} \}}{} +F\cdot M_u -M\label{CAseq:SC3}\\
&=&\expectedbig{\sum_{j=1}^b \sum_{f=1}^F S_f \big ( \mathbb{I} \{f \notin {\mathcal{M}}^{n_{j-1}}, f \in {\mathcal{M}}^{n_j}, {\mathcal{M}}^{n_j} \in \mathcal{B} \}  \nonumber \\ 
&& +   \mathbb{I} \{f \notin{\mathcal{M}}^{n_{j-1}}, f \in {\mathcal{M}}^{n_j}, {\mathcal{M}}^{n_j} \notin \mathcal{B} \} \big ) }{}+F\cdot M_u -M\label{CAseq:SC4}\\
&=&\expectedbig{\sum_{j=1}^b \sum_{f=1}^F S_f \big ( \mathbb{I} \{T_{f,n_j} > T_{f,n_{j}-1}, f \notin {\mathcal{M}}^{n_{j-1}}, {\mathcal{M}}^{n_j} \in \mathcal{B} \} \nonumber \\ 
&& +   \mathbb{I}  \{f \notin {\mathcal{M}}^{n_{j-1}}, f \in{\mathcal{M}}^{n_j}, {\mathcal{M}}^{n_j} \notin \mathcal{B} \} \big )}{} +F\cdot M_u -M\label{CAseq:SC5}\\
&\leq & M_u \expected{ \sum_{j=1}^b  \mathbb{I} \left \{ \sum_{f=1}^F  T_{f,n_j} > \sum_{f=1}^F  T_{f,n_{j}-1}, f \notin {\mathcal{M}}^{n_{j-1}}, {\mathcal{M}}^{n_j} \in \mathcal{B} \right \} }{} \nonumber\\ 
&& + \expected{\sum_{j=1}^b\sum_{f=1}^F   \mathbb{I}  \{f \notin {\mathcal{M}}^{n_{j-1}}, f \in{\mathcal{M}}^{n_j}, {\mathcal{M}}^{n_j} \notin \mathcal{B} \}  }{} \!+\!F\!\cdot\!M_u\!-\!M\label{CAseq:SC6}\\
&\leq&  M_u\expected{ \sum_{j=1}^b \mathbb{I} \left \{\sum_{f=1}^F N_{f,n_j} > \sum_{f=1}^F N_{f,n_{j}-1} \right \} }{} \nonumber \\
&&\!+\!\expected{  \sum_{j=1}^b \sum_{f=1}^F  S_f \mathbb{I} \{f\!\notin\!{\mathcal{M}}^{n_{j-1}}, f \in {\mathcal{M}}^{n_j},{\mathcal{M}}^{n_j}\!\notin\!\mathcal{B} \}}{}\!+\!F\!\cdot\!M_u\!-\!M\label{CAseq:SC8}\\
&=& M_u\underbrace{ \sum_{j=1}^b\frac{ \overline{N}_{n_{j+1}-1} -\overline{N}_{n_{j}-1}}{\Delta (j)} }_{Sw^\pi_{\mathcal{B}}(t)} +  \expected{  \sum_{j=1}^b \sum_{f=1}^F  S_f \mathbb{I} \{f\!\notin\!{\mathcal{M}}^{n_{j-1}}, f \in {\mathcal{M}}^{n_j}, {\mathcal{M}}^{n_j}\!\notin\!\mathcal{B} \}}{}\!\!+\!F\!\cdot\!M_u\!-\!M\label{CAseq:SC9}\\
&\leq& M_u Sw^\pi_{\mathcal{B}}(t)\!+\!\left(1\!+\!Sw^\pi_{\mathcal{B}}(t)\right)M_u\!+\!\left(b\!-\!2 Sw^\pi_{\mathcal{B}}(t)\!-\!1\right) M_{l}\!+\!F\!\cdot\!M_u\!-\!M\label{CAseq:SC10}\\
&\leq&Sw^\pi_{\mathcal{B}}(t) \cdot 2(M_u-M_{l}) + (b-1)\cdot M_{l} +F\cdot M_u,\label{CAseq:SC11}
\end{IEEEeqnarray}
\end{subequations}
where \eqref{CAseq:SC3} holds since in the algorithm initalization the switching cost is maximum; in the  first line of \eqref{CAseq:SC4}  we split the switching periods into those we switch to a good arm combination (i.e., $\mathcal{M}^{n_j}\notin \mathcal{B}$), and those we switch to a bad arm combination (i.e., $\mathcal{M}^{n_j}\in \mathcal{B}$); in \eqref{CAseq:SC5} we use the fact that $T_{f,t}$ is updated each time an arm is played;  \eqref{CAseq:SC6} follows since the cost of a bad is bounded by $M_u$; \eqref{CAseq:SC8} follows since only one counter $N_{f,n}$ is updated per bad period, and we sum over all arms instead of summing over the arms that were not played in the last period, in \eqref{CAseq:SC9} we use the fact that $\frac{ \sum N_{f,n_{j+1}-1} -\sum N_{f,n_{j}-1}}{\Delta (j)}$ is one if $\sum N_{f,n_{j+1}} > \sum N_{f,n_{j}}$, and zero otherwise; the bound in  \eqref{CAseq:SC10}  is obtained by assuming that after each bad period a good arm combination is played (incurring a cost of $M_u$), and hence, the number of consecutive plays of good arm combinations is minimised, finally in \eqref{CAseq:SC11} we use $M_u \leq M$. This proves  \eqref{CAeq:Sw}.  

We note that $Sw^{\pi}_{\mathcal{B}}(t)$, which represents a bound on the number of switches between bad arm combinations can be rewritten as
\begin{IEEEeqnarray}[\eqnsize]{rCL}\label{CAeq:A:SwB}
 Sw^\pi_{\mathcal{B}}(t)
 &=&\frac{\overline{N}_{n_{b+1}-1}}{\Delta (b)} - \frac{F}{\Delta (1)}+\sum_{j=2}^{b} \overline{N}_{n_{j}-1}  \nabla(j),
\end{IEEEeqnarray}
where, $\nabla(j)=\frac{1}{\Delta (j-1)} -\frac{1}{\Delta (j)}$. This expression is used in the rest of the proof.

\subsection{Proof of Theorem~\ref{CAthm:sqrt}} \label{CAapp:proof_sqrt}
We begin by stating two properties that are used later in the proof of Theorem~\ref{CAthm:sqrt}.

\begin{property} \label{CAprop1} If $\Delta(j)=\left \lceil \gamma \sqrt{n_j}\right \rceil$ and $ \gamma \geq 2+\frac{1}{\sqrt{F+1}}$, we have  $\frac{\Delta (j)}{n_j} \leq \frac{\Delta (j-1)}{n_{j-1}}$.
\end{property}
\begin{property} \label{CAprop2} If $\Delta(j)=\left \lceil \gamma \sqrt{n_j}\right \rceil$ and $ \gamma \geq 2+\frac{1}{\sqrt{F+1}}$ we have  $b\leq\sqrt{n_b}$.
\end{property}
Properties~\ref{CAprop1} and~\ref{CAprop2} are proven in Appendixes \ref{CAapp:prop1} and~\ref{CAapp:prop2}, respectively.

Moreover, $\Delta(j)$ is a monotonically increasing function of $j$; and hence, $\displaystyle \max_{1\leq i\leq j} \Delta(i)=\Delta(j)$, and $\gamma \sqrt{n_i} \leq \Delta(i) \leq \gamma \sqrt{n_i} +1 $. 
\subsubsection{Sampling-regret}
Using  $\Delta(j)\leq \gamma \sqrt{n_j}+1$ in (\ref{CAeq:A:sampling-regret}), we obtain 
\begin{equation}\label{CAeq:A:sampling-regret-sqrt}
R^{\pi}_{Sa}(t) \leq \left(\frac{\pi^2}{3}+4.12\gamma+l_t+ \gamma \sqrt{t}+1 \right) F  \Delta_{u},
\end{equation}
where we have used that $K_1 = \frac{\pi^2}{3}+4.12\gamma$, which is obtained as the sum of two Riemann zeta functions with parameters $2$ and $\frac{3}{2}$.
\subsubsection{Switching-regret}
Substituting (\ref{CAeq:A:EN_bound}) into (\ref{CAeq:A:SwB}), we get 
\begin{subequations}\label{CAeq:A:SwB_bound}
\begin{IEEEeqnarray}[\small]{rcl}
 Sw^\pi_{\mathcal{B}}(t) 
 &\leq& F\left(\frac{l_{n_{b+1}-1}}{\Delta(b)}+\sum_{j=2}^{b} l_{n_j-1}\nabla(j)\right)+(1-\beta) \left(\frac{n_{b+1}-1-F}{\Delta(b)}+\sum_{j=2}^{b} (n_j-1-F) \nabla(j)\right) \nonumber \\
 &&-\frac{F}{\Delta(1)}+F\cdot K_1\left(\frac{1}{\Delta(b)}+\sum_{j=2}^{b}\nabla(j)\right) +F\left(\frac{\Delta(b-1)}{\Delta(b)}+\sum_{j=2}^{b}\Delta(j-1)\nabla(j)\right)  \label{CAseq:swr2}\\
 &=& F    \left(\frac{l_{n_{b+1}-1}}{\Delta(b)} + \sum_{j=2}^{b} l_{n_j-1}\nabla(j)\right)   + (1-\beta)\left ( \frac{n_{b+1}-1-F}{\Delta(b)} + \frac{n_2 -1 -F}{\Delta(1)}-\frac{n_b -1 -F}{\Delta(b)} +b-2\right) \nonumber \\
  &&-\frac{F}{\Delta(1)}+ \frac{F\cdot K_1}{\Delta(2)}+ F   \left(\frac{\Delta(b-1) }{\Delta(b)} + \sum_{j=2}^{b}  \Delta(j-1)\nabla(j)\right) \label{CAseq:swr3} \\  
 &\leq&F   \left(\frac{l_{n_{b+1}-1}}{\Delta(b)} + \sum_{j=2}^{b} l_{n_j-1}\nabla(j)\right)  + (1-\beta)\left ( \frac{n_1}{\Delta(1)} -\frac{1+F}{\Delta(1)} +b \right)  \nonumber\\
 &&-\frac{F}{\Delta(1)}+ \frac{F\cdot K_1}{\Delta(2)} + F   \left(1+\sum_{j=2}^b \left ( \frac{\Delta(j)}{\Delta(j-1)}-\frac{\Delta(j)}{\Delta(j)}\right) \right) \label{CAseq:swr4}\\
  &\leq&F   \left(\frac{l_{n_{b+1}}}{\Delta(b)} + \sum_{j=2}^{b} l_{n_j}\nabla(j)\right) + (1-\beta)\sqrt{t} -\frac{F}{\Delta(1)}+ \frac{F\cdot K_1}{\Delta(2)}  + F   \left(1+\sum_{j=2}^b \left ( \frac{n_j}{n_{j-1}}-1 \right)\right)  \label{CAseq:swr6}
\end{IEEEeqnarray}
\end{subequations}
where \eqref{CAseq:swr2} follows from \eqref{CAeq:A:EN_bound} and \eqref{CAeq:A:SwB}; \eqref{CAseq:swr3} follows since  $\sum_{j=2}^{b} (n_j-1-F) \nabla(j)$ and 
$\sum_{j=2}^{b}\nabla(j)$ in the first and second lines of \eqref{CAseq:swr2} are the sum of telescopic minus a constant series and a telescopic series, respectively;    \eqref{CAseq:swr4} follows from the fact that  $\Delta(j)$ is monotonically increasing, and the definition of $\nabla(j)$; \eqref{CAseq:swr6} holds since $n_1=F+1$, $l_t$ is monotonically increasing, and Property~\ref{CAprop2} applies to the first line of \eqref{CAseq:swr4} and Property~\ref{CAprop1} applies to the second line  of \eqref{CAseq:swr4}, and finally $n_b \leq t \leq n_{b+1}$. 
We  study separately the first and last sums in  \eqref{CAseq:swr6}. For the first sum we have  
\begin{subequations}
\begin{IEEEeqnarray}[\eqnsize]{rCL}
\frac{l_{n_{b+1}}}{\Delta(b)} + \sum_{j=2}^{b} l_{n_j}\nabla(j)
 &=& \frac{6}{(g^{-1}(\Delta_{l}))^2} \left ( \frac{\log(n_1)}{\Delta (1)} +  \sum_{j=1}^{b} \frac{\log(n_{j+1})-\log(n_j)}{\Delta (j)} \right ) \label{CAseq:swrK23} \\
 &=& \frac{6}{(g^{-1}(\Delta_{l}))^2} \left ( \frac{\log(n_1)}{\Delta (1)} +  \sum_{j=1}^{b} \frac{\log\left (1+\frac{\Delta(j)}{n_j} \right )}{\Delta (j)} \right ) \label{CAseq:swrK24}\\
 &\leq&  \frac{6}{(g^{-1}(\Delta_{l}))^2} \left ( \frac{\log(n_1)}{\Delta (1)} +  \sum_{j=1}^{b} \log\left (1+\frac{\Delta(j)}{n_j} \right ) \right ) \label{CAseq:swrK25}\\
 &\leq&  \frac{6}{(g^{-1}(\Delta_{l}))^2} \left ( \frac{\log(F+1)}{\lceil \gamma \sqrt{F+1} \rceil }+\log(n_b)\right ), \label{CAseq:swrK26} 
\end{IEEEeqnarray}
\end{subequations}
where in \eqref{CAseq:swrK23} we use the definition of $l_t$ and $\nabla(j)$, and the fact that the summation is a telescopic sum; in 
\eqref{CAseq:swrK25} that $\Delta(j) \geq 0$;  and the last inequality is proven by induction in Appendix~\ref{CAapp:prop3} and holds for $\gamma \leq \frac{F^2+F-1}{\sqrt{F+1}}$.  For the second summation in \eqref{CAseq:swr6} we have  
 \begin{subequations}
\begin{IEEEeqnarray}[\eqnsize]{rCL}
  \sum_{j=2}^b\left ( \frac{n_j}{n_{j-1}} -1\right) 
 &=&  \sum_{j=2}^b\frac{\Delta(j-1) }{n_{j-1}} \\
 &\leq&  \sum_{j=2}^b \frac{\gamma \sqrt{n_{j-1}}+1}{n_{j-1}}\\
 &\leq&  \sum_{j=2}^b \left ( \frac{\gamma}{j} +\frac{1}{j^2} \right ) \label{CAseq:k3_3}\\
 &\leq & \gamma\left (1+\log(b)\right)+\frac{\pi^2}{6} \label{CAseq:k3_4}\\
 &\leq & \gamma\left (1+\frac{1}{2}\log(t)\right ) +\frac{\pi^2}{6}, \label{CAseq:k3_5}
  \end{IEEEeqnarray}
\end{subequations}
where in \eqref{CAseq:k3_3} we have used Property \ref{CAprop2}, in \eqref{CAseq:k3_4} we have applied the divergence rate and the infinite sum of the harmonic series, and in \eqref{CAseq:k3_5} we have used Property~\ref{CAprop2}.

Using the bounds in \eqref{CAseq:swrK26} and \eqref{CAseq:k3_5} and applying $K_1 = \frac{\pi^2}{3}+4.12\gamma$ into (\ref{CAseq:swr6}) we get
 \begin{IEEEeqnarray}[\eqnsize]{rCL}\label{CAeq:A:SwB_bound2}
 Sw^{\pi}_{\mathcal{B}}(t) &\leq&    \frac{6F}{(g^{-1}(\Delta_{l}))^2} \left ( \frac{\log(F+1)}{\lceil \gamma \sqrt{F+1} \rceil } +  \log(t) \right ) + (1-\beta)\sqrt{t} \nonumber\\
 &&+\!F\!\left(\!1\!+\!\frac{\pi^2}{6}\!+\!\gamma\left(1\!+\!\frac{1}{2}\log(t)\right)\!\right)\!+\!F\frac{\frac{\pi^2}{3}+4.12\cdot\gamma}{\Delta(2)} -\frac{F}{\Delta(1)},
  \end{IEEEeqnarray}
  Finally, from (\ref{CAeq:A:SwB_bound2}) and (\ref{CAeq:Sw}) we get 
\begin{IEEEeqnarray}[\eqnsize]{rCL} \label{CAeq:A:Sw_bound3} 
R_{Sw}^{\pi} (t)&\leq& 
\sqrt{t} \left[ M_u\!+\!(M_u-M_l)(1\!-\!2\beta) \right] + \log{t} \left(  \frac{6}{(g^{-1}(\Delta_{l}))^2} + \frac{\gamma}{2}\right)\cdot 2 F(M_u-M_l)\!+\!C,
\end{IEEEeqnarray}
where $C=F\left ( \frac{6F}{(g^{-1}(\Delta_{l}))^2} \cdot \frac{\log(F+1)}{\lceil \gamma \sqrt{F+1} \rceil } 1+\frac{\pi^2}{6}+\gamma + \frac{\frac{\pi^2}{3}+4.12\cdot\gamma}{\Delta(2)} -\frac{1}{\Delta(1)} \right)\cdot 2 F(M_u-M_l)+FM_u-M_l$ is a constant that does not depend on $t$. 

\subsubsection{Regret with switching cost}
From (\ref{CAeq:A:Sw_bound3}), (\ref{CAeq:A:sampling-regret-sqrt}), and (\ref{CAeq:regret}),  we get  the regret bound of Theorem~\ref{CAthm:sqrt}.  We note that $\forall F \geq 2$ there exist a $\gamma$ such that $\frac{F^2+F-1}{\sqrt{F+1}}\geq \gamma \geq 2+\frac{1}{\sqrt{F+1}}$.

\subsection{Proof of Theorem~\ref{CAthm:L}} \label{CAapp:proof_L}
We consider the CUCBSC algorithm with $\Delta(j)=L$, $\forall j$.  That is, $n_b=F+1+L(b-1)$, which implies that $\nabla(j)=0$, $\forall j$, and $\max_{1\leq j\leq b}\Delta(j)=L$, $\forall b$.
\subsubsection{Sampling-regret}
Using  $\Delta(j)=L$ in (\ref{CAeq:A:sampling-regret}), we get  
\begin{equation}\label{CAeq:A:sampling-regret-L}
R^{\pi}_{Sa}(t) \leq \left(\frac{\pi^2}{3L}+l_t+ L \right) F  \Delta_{u},
\end{equation}
where, $K_1=\frac{\pi^2}{3L}$ is obtained from the Riemann zeta function with parameter~$2$.
\subsubsection{Switching-regret}
 Using \eqref{CAeq:A:EN_bound} and the fact that, for $\Delta(j)=L$,  $\nabla(j)=0$ in (\ref{CAeq:A:SwB}), we obtain
\begin{IEEEeqnarray}[\eqnsize]{rCL}  \label{CAeq:A:SwB_bound_linear}
Sw^\pi_{\mathcal{B}}(t)  &\leq&  
 \frac{F}{L} \left (\frac{\pi^2}{3L}+\frac{6\log t+\log \left(1+\frac{L-1}{F+1} \right)}{(g^{-1}(\Delta_{l}))^2}+ L -1\right),
\end{IEEEeqnarray}
 where we have also used $l_t=\frac{6\log t}{(g^{-1}(\Delta_{l}))^2}$, $\beta=1$, and that $n_{b+1} \leq t+L$. To obtain the sampling regret we use the fact that the \mbox{($\alpha,\beta$)-solver}  always finds the unique optimal solution, and hence, $\beta=1$ and $M_l=0$, and plug (\ref{CAeq:A:SwB_bound_linear}) into~(\ref{CAeq:A:Sw}). 
\subsubsection{Regret with switching cost}
We obtain the results of Theorem~\ref{CAthm:L} by plugging into (\ref{CAeq:regret}) the switching regret obtained from (\ref{CAeq:A:sampling-regret-L}), and (\ref{CAeq:A:SwB_bound_linear}). 
\section{Proof of Property \ref{CAprop1}}\label{CAapp:prop1}
Property~\ref{CAprop1} can be expressed as
\begin{equation} \label{CAeq:prop1_proof1}
\Delta (j) \leq   \Delta (j-1)  + \frac{\Delta(j-1)^2}{n_{j-1}}.
\end{equation}
Using   $\Delta(j) \leq \gamma \sqrt{n_{j-1} + \Delta (j-1) } +1$, we can show that if 
\begin{equation} \label{CAeq:prop1_proof2}
\gamma \sqrt{n_{j-1} + \Delta (j-1) } \leq  \Delta (j-1)  + \frac{\Delta(j-1)^2}{n_{j-1}} -1,
\end{equation}
then \eqref{CAeq:prop1_proof1} holds. Both sides of \eqref{CAeq:prop1_proof2} are positive, taking the square power, we obtain
\begin{IEEEeqnarray}[\eqnsize]{rcl}\label{CAeq:prop1_proof3}
\gamma^2 (n_{j-1} \!+\! \Delta(j\!-\!1) )&\leq& \Delta(j\!-\!1)^2 \!\!+\!\! \frac{\Delta(j-1)^4}{{n_{j-1}}^2}\!-\!2\frac{\Delta(j-1)^2}{n_{j-1}}\!+\!2\frac{\Delta(j\!-\!1)^3}{n_{j-1}}\!-\!2\!\Delta(j-1) +1.
\end{IEEEeqnarray}
From the left-hand side (LHS) and right-hand side (RHS) of \eqref{CAeq:prop1_proof3} we get 
\begin{IEEEeqnarray}[\eqnsize]{rCL}
\text{ \{LHS of\} \eqref{CAeq:prop1_proof3}}&\leq & \gamma^2 (n_{j-1} +1) + \gamma^3\sqrt{n_{j-1}} \label{CAeq:prop1_proof4} \\ 
\text{\{RHS of\} \eqref{CAeq:prop1_proof3}}&\geq& \gamma^4+\sqrt{n_{j-1}}2\gamma(\gamma^2-1)+\gamma^2(n_{j-1}-2)-\frac{4\gamma}{\sqrt{n_{j-1}}}-\frac{2}{n_{j-1}}-1 \label{CAeq:prop1_proof5},
\end{IEEEeqnarray}
%
where \eqref{CAeq:prop1_proof5} follows since $-\Delta(j-1) \geq -\gamma \sqrt{n_{j-1} } -1$, and  $\Delta(j-1) \geq \gamma \sqrt{n_{j-1} }$. Subtracting  \eqref{CAeq:prop1_proof4} from \eqref{CAeq:prop1_proof5} we get 
\begin{subequations}
\begin{IEEEeqnarray}[\eqnsize]{rCL}\label{CAeq:prop1_proof6}
\text{ \eqref{CAeq:prop1_proof5} $-$ \eqref{CAeq:prop1_proof4}}&=& \gamma^4+\sqrt{n_{j-1}}\gamma\left(\gamma^2-2-\frac{4}{n_{j-1}}\right)-3\gamma^2-\frac{2}{n_{j-1}}-1  \\
&&\geq\gamma^4-3\gamma^2-\frac{2}{n_{j-1}}-1  \label{CAseq:prop1_pol1}\\
&&\geq \gamma^4-3\gamma^2-\frac{2}{F+1}-1  \label{CAseq:prop1_pol2} 
\end{IEEEeqnarray}
\end{subequations}
where in \eqref{CAseq:prop1_pol1} we have used that since $\gamma \geq 2 +\frac{1}{\sqrt{1+F}}$, we have $\gamma \geq \sqrt{2+\frac{4}{F+1}}$ for $F\geq2$; and in \eqref{CAseq:prop1_pol2} that $n_j\geq F+1$. Note that \eqref{CAseq:prop1_pol2} is a polynomial of order four with a single real root. Since we assume that $\gamma\geq 2 +\frac{1}{\sqrt{1+F}}$, $\gamma$  is always greater than $\frac{\sqrt{13+\frac{8}{F+1}}-3}{2}$ and \eqref{CAseq:prop1_pol2} is positive. This proves that \eqref{CAeq:prop1_proof3} holds, and the proof is completed.   

\section{Proof of Property \ref{CAprop2}}\label{CAapp:prop2}
This is proven by induction. For $b=1$ it is clear since $n_1=F+1$. We have
\begin{subequations}\label{CAseq:prop1}
\begin{IEEEeqnarray}[\eqnsize]{rCL} 
\sqrt{n_{b+1}}-1 
&\geq& \sqrt{n_{b}+ \gamma \sqrt{n_b} }-1 \label{CAseq:prop1_1} \\
&=& \sqrt{n_{b}} \left( \sqrt{1+\frac{\gamma}{\sqrt{n_b}}} -\frac{1}{\sqrt{n_b}}\right)\label{CAseq:prop1_2}  \\
&\geq& b \left( \sqrt{1+\frac{\gamma}{\sqrt{n_b}}} -\frac{1}{\sqrt{n_b}}\right) \label{CAseq:prop1_3}  \\
&\geq& b, \label{CAseq:prop1_4} 
\end{IEEEeqnarray}
\end{subequations}
where \eqref{CAseq:prop1_3} follows from the induction hypothesis; \eqref{CAseq:prop1_4} holds since $\gamma \geq 2+\frac{1}{\sqrt{F+1}}$ ensures that $\gamma \geq 2+\frac{1}{\sqrt{n_b}}$.  

\section{}\label{CAapp:prop3}
We need to show that  $\sum_{j=1}^{b} \log\left (1+\frac{\Delta(j)}{n_j} \right ) \leq \log(n_b), \forall b$. We use induction again. For $b=1$, the inequality  holds if $\gamma \leq \frac{F^2+F-1}{\sqrt{F+1}}$. 
We also have
 \begin{subequations}
\begin{IEEEeqnarray}[\eqnsize]{rCL}
\sum_{j=1}^{b+1} \log\left (1+\frac{\Delta(j)}{n_j} \right ) &=& \sum_{j=1}^{b} \log\left (1+\frac{\Delta(j)}{n_j} \right ) + \log\left (1+\frac{\Delta(b+1)}{n_{b+1}} \right ) \label{CAseq:app3ih1}\\
&\leq& \log(n_{b})  + \log\left (1+\frac{\Delta(b+1)}{n_{b+1}} \right ) \label{CAseq:app3ih2}\\
&\leq& \log \left (n_{b}  + \frac{n_b \Delta (b)}{n_{b}} \right ) \label{CAseq:app3ih4}\\
&= & \log(n_{b+1}), \label{CAseq:app3ih6}
\end{IEEEeqnarray}
\end{subequations}
where we have used the induction hypothesis in \eqref{CAseq:app3ih2},and  Property~\ref{CAprop1} in \eqref{CAseq:app3ih4}.

\begin{spacing}{1.4}
\bibliographystyle{IEEEtran}
\bibliography{IEEEabrv,Totabiblio}
\end{spacing}

\end{document}